\documentclass[journal,draftclsnofoot,onecolumn,12pt]{IEEEtran}

\usepackage{amsthm,amssymb,graphicx,multirow,amsmath,color,amsfonts}
\usepackage[update,prepend]{epstopdf}
\usepackage[noadjust]{cite}
\usepackage[latin1]{inputenc}
\usepackage{tikz}
\usetikzlibrary{arrows,calc}		
\usepackage{bbm} 
\usepackage{pdfpages}
\usepackage{tabulary}
\usepackage{multirow}
\usepackage{comment}
\usepackage{dsfont}
\usepackage{pgf,tikz}
\usepackage{mathrsfs}
\usetikzlibrary{arrows}
\usepackage{subcaption}
\usepackage{bigints}
\usepackage{mathtools}

\usepackage{setspace}	
\newcommand{\bsloc}[1]{{\bf b}_{#1}}
\newcommand{\ueloc}[1]{{\bf u}_{#1}}

\newcommand{\lt}{\mathtt{LT}}

\newcommand{\cdf}{\mathtt{CDF}}
\newcommand{\pdf}{\mathtt{PDF}}

\newcommand\numberthis{\addtocounter{equation}{1}\tag{\theequation}}
\newcommand{\varn}[1]{\mathtt{Var}\big[#1\big]}

\newcommand{\numUE}{N}

\newcommand{\setCCP}{{\cal C}}

\newcommand{\setCEP}{{\cal E}}
\newcommand{\vCell}{{\cal V}}
\newcommand{\ccReg}{{\cal X}_C}
\newcommand{\ccAre}{{X}_C}
\newcommand{\ceAre}{{X}_E}
\newcommand{\ccPhi}{{\Phi}_{\tt u, k}^{\mathrm{PPP}}}
\newcommand{\ccPhiDen}{{\lambda}_{\tt u, k}^{\mathrm{PPP}}}
\newcommand{\ccPhiIM}{{\Lambda}_{\tt u, k}^{\mathrm{PPP}}}

\newcommand{\cePhi}{{\Phi}_{\tt u, l}^{\mathrm{PPP}}}
\newcommand{\cePhiDen}{{\lambda}_{\tt u, l}^{\mathrm{PPP}}}
\newcommand{\cePhiIM}{{\Lambda}_{\tt u, l}^{\mathrm{PPP}}}

\def\nbb{{\mathbf{b}}}

\def\nbg{{\mathbf{g}}}
\def\nbh{{\mathbf{h}}}

\def\nbn{{\mathbf{n}}}
\def\nbo{{\mathbf{o}}}
\def\nbp{{\mathbf{p}}}

\def\nbr{{\mathbf{r}}}

\def\nbu{{\mathbf{u}}}
\def\nbv{{\mathbf{v}}}
\def\nbw{{\mathbf{w}}}
\def\nbx{{\mathbf{x}}}
\def\nby{{\mathbf{y}}}

\def\nb0{{\mathbf{0}}}
\def\nb1{{\mathbf{1}}}

\def\nbI{{\mathbf{I}}}

\def\nbbC{{\mathbb{C}}}

\def\nbbE{{\mathbb{E}}}

\def\nbbR{{\mathbb{R}}}

\newtheorem{lemma}{Lemma}

\newtheorem{prop}{Proposition}

\newtheorem{remark}{Remark}

\def\R{\mathbb{R}}

\def\sinr{\mathtt{SINR}}			

\def\se{\mathtt{SE}}
\def\cse{\mathtt{CSE}}

\newcommand{\dP}[1]{\mathbb{P}\left[#1\right]}
\newcommand{\dE}[2]{\mathbb{E}_{#2}\left[#1\right]}

\allowdisplaybreaks 

\begin{document}
\bstctlcite{IEEEexample:BSTcontrol}

\graphicspath{{./Figures/}}
\title{Stochastic Geometry-based Uplink Analysis of Massive MIMO Systems with Fractional Pilot Reuse}

\author{
Priyabrata Parida and Harpreet S. Dhillon. \vspace{-1cm}
\thanks{The authors are with Wireless@VT, Department of ECE, Virginia Tech, Blacksburg, VA. Email: \{pparida, hdhillon\}@vt.edu. The support of the US NSF (Grant ECCS-1731711) is gratefully acknowledged. This paper will be presented in part at the IEEE International Conference in Communications (ICC), 2018~\cite{Parida2018}. \hfill}
}

\maketitle
\vspace{-0.8cm}
\begin{abstract}
In this work, we analyze the performance of the uplink (UL) of a massive MIMO network considering an asymptotically large number of antennas at base stations (BSs). 
We model the locations of BSs as a homogeneous Poisson point process (PPP) and assume that their service regions are limited to their respective Poisson-Voronoi cells (PVCs).
Further, for each PVC, based on a threshold radius, we model the {\em cell center} (CC) region as the {\em Johnson-Mehl} (JM) cell of its BS while rest of the PVC is deemed as the {\em cell edge} (CE) region. 
The CC and CE users are located uniformly at random independently of each other in the JM cell and CE region, respectively.
In addition, we consider a fractional pilot reuse (FPR) scheme where two different sets of pilot sequences are used for CC and CE users with the objective of reducing the interference due to pilot contamination for CE users.
Based on the above system model, we derive analytical expressions for the UL signal-to-interference-and-noise ratio ($\sinr)$ coverage probability and average spectral efficiency (SE) for randomly selected CC and CE users.
In addition, we present an approximate expression for the average cell SE.
One of the key intermediate results in our analysis is the approximate but accurate characterization of the distributions of the CC and CE areas of a typical cell. Another key intermediate step is the accurate characterization of the pair correlation functions of the point processes formed by the interfering CC and CE users that subsequently enables the coverage probability analysis. 
From our system analysis, we present a partitioning rule for the number of pilot sequences to be used for CC and CE users as a function of threshold radius that improves the average CE user SE while achieving similar CC user SE with respect to unity pilot reuse.
\end{abstract}


\section{Introduction}
Owing to its ability to improve both spectral and energy efficiency of wireless networks, massive multiple-input multiple-output (mMIMO) is considered a key enabler of the fifth-generation (5G) communication systems and beyond. 
Fundamentally, mMIMO is a multi-user MIMO system where a large number of antennas at the base stations (BSs) are used to simultaneously serve a fewer number of users (compared to  the number of antennas at the BSs).
Although a simple extension of conventional multi-user MIMO technique, it is set to revolutionalize wireless communication networks as it has been proven that under ideal conditions it eliminates the deleterious effect of channel fading and additive noise while negating the effect of network interference~\cite{Mar2010, Emil2016, Emil2017}.
In order to decode the simultaneously transmitted data from different users, each BS requires the channel knowledge of the users attached to it that is estimated through a set of orthogonal pilot sequences.
Due to limited channel coherence interval, the number of orthogonal pilot sequences is also limited.
As a result, the pilot sequences need to be reused across different cells. 
In his seminal work~\cite{Mar2010}, Marzetta showed that under the assumption of independent and identically distributed (i.i.d.) Rayleigh fading across BS antennas and sub-optimal low-complexity processing schemes such as maximal ratio combining (MRC), the reuse of pilot sequences gives rise to an inherent interference known as {\em pilot contamination} (PC), which fundamentally limits the performance of mMIMO networks.
As discussed next in detail, a significant amount of research effort has been focused on overcoming the effect of PC. 
Amongst all the solutions, a relatively simple scheme, namely fractional pilot reuse (FPR), stands out in reducing the effect of PC, especially for the cell edge (CE) users.
{The objective of this article is to analyze the performance of a mMIMO network that uses the FPR scheme.}

\subsection{Motivation and related works}
In the literature, different methods that aim to suppress or mitigate the effect of PC can be broadly categorized into protocol based methods~\cite{Fern2013}, BS coordination based methods~\cite{Yin2013, Jose2011}, and pilot reuse or hopping based methods~\cite{SorCar2014, Alwa2017}. 
Please refer to~\cite{Elijah2016} for a comprehensive survey on this subject.
While protocol and coordination based methods are effective in removing the PC, they are usually complex. 
Further, these techniques require some form of coordination among BSs to obtain reliable channel estimates.
On the other hand, the gains obtained by pilot hopping based methods is primarily due to interference randomization and is hence limited to scenarios with larger channel coherence times.
In contrast, a low complexity and distributed scheme to counter the effect of PC is to forbid reusing the same pilots  in every cell, which requires limited synchronization and no coordination among BSs~\cite{Li2012, Bjor2016}. 
The concept of pilot reuse is similar to the frequency reuse in cellular networks.
In~\cite{Li2012}, the optimal pilot reuse factor is obtained for a network with linear topology. From the numerical simulations, authors show that higher than unity pilot reuse factor is beneficial for average cell throughput.
In~\cite{Bjor2016}, for a hexagonal cellular network model, authors show that unity pilot reuse may not be optimal in all scenarios.
Above-mentioned works focused on the scenario where orthogonal sets of pilots are used in neighboring cells.
However, the spectral efficiency (SE) can be further improved by using a more aggressive pilot reuse scheme, namely FPR, instead of completely orthogonal reuse across cells.
Conceptually, FPR is similar to that of fractional frequency reuse (FFR) used in LTE systems to mitigate the effect of inter-cell interference. 
To the best of the knowledge of the authors, the concept of FPR was first introduced in~\cite{Atzeni2015}.
{In FPR, similar to FFR, depending on the channel condition, users in a cell are classified into two categories, namely cell center (CC) and  CE users.} 
While the set of pilots reserved for CC users are reused in every cell, the set of pilots for CE users are reused in specific cells depending upon the reuse factor.
In contrast to the sophisticated and complex coordination schemes discussed earlier, FPR requires no coordination among BSs and is hence simpler to implement in practice.

For the performance analysis of mMIMO systems with FPR, it is imperative to consider a large-scale multi-cell setup so that the effect of interference on the performance can be accurately modeled. For such problems, stochastic geometry provides a rigorous set of tools for the spatial modeling and performance analysis, as discussed in detail in~\cite{AndGupJ2016, ElSawy2017}. 
For a pedagogical treatment of the subject with emphasis on the application to cellular network, interested readers are advised to refer to~\cite{AndGupJ2016}.
Although stochastic geometry has been used for the performance analysis of mMIMO systems in \cite{Madhu2013, Govindasamy2014, BaiH2014, BaiHea2015B, Govindasamy2015, BaiH15, BjornsonSK15a}, the analyses presented in these works cannot be trivially extended to accommodate the analysis of FPR scheme.
One important reason behind this is that the UL interference field generated by the users in FPR scheme is different from unity pilot reuse scheme, which is considered in the above-mentioned works.
Further, the analyses (except in ~\cite{Madhu2013}) are limited to the consideration of a fixed number of users in the network which does not take into account the load (number of active users) attached with BSs.
In this work, we propose a new approach to analyze the performance of a mMIMO network considering FPR scheme that results in the following key contributions.

\subsection{Contributions of the work}
\subsubsection{Analytical model for UL analysis of a mMIMO system with FPR}
A new generative model is proposed to analyze the performance of the UL of a mMIMO system in the asymptotic antenna regime under the consideration of FPR scheme. 
We model the BS locations as a Poisson point process (PPP). Based on a threshold distance $R_c$, we characterize the CC regions as the Johnson-Mehl (JM) cells associated with the BSs. The complementary region in each cell is modeled as the CE region.
One important result in our analyses is the approximate but accurate distribution functions for the CC and CE areas of a typical Poisson-Voronoi Cell (PVC). These results are subsequently used to model the load (number of CC and CE users) distribution of each cell.
{Using these distributions, we provide key intermediate results, such as the pilot assignment probability of a randomly selected CC (CE) user and utilization probability of a pilot. These results are later used in the coverage probability and SE analyses.}

\subsubsection{Signal-to-interference-plus-noise ratio ($\sinr$) coverage, average user and cell SEs analysis}
We present $\sinr$ coverage probability of a user assigned to a given CC (CE) pilot.  
The derivation of exact probability is difficult as the exact statistical characterization of the interference field is extremely challenging.
In fact, derivation of this result for a relatively simpler scenario of the classical UL system, where the segregation between CC and CE users is not present, is also intractable.
Hence, to lend tractability to this problem, we resort to a careful approximation of the interference statistics in the UL. 
Motivated by~\cite{Haenggi2017}, first, we derive the pair correlation function (PCF) of the interfering user locations with respect to (w.r.t.) the BS of interest. Using this PCF, we approximate the point process formed by the CC (CE) interfering users as a non-homogeneous PPP.
{Next, based on the {\em dominant interferer} based approach, we provide useful theoretical expressions for the coverage probability of a user assigned to a CC (CE) pilot. This result is extended to obtain analytical expressions for the average SEs of a randomly selected CC (CE) user and average SE of a typical cell.}

\subsubsection{System design guidelines}
Our analysis leads to following system design guidelines.
{First, our analyses show that for a certain range of threshold radius, by allocating $1-\exp(-\pi \lambda_0 c_2 R_c^2)$, where $\lambda_0$ is the BS density and $c_2$ is a constant,  fraction of pilots for the CC users, FPR scheme improves the average SE of a CE user without affecting the average SE of a CC user compared to unity reuse.}
Second, for a given threshold radius, it is possible to achieve higher average cell SE using FPR scheme compared to unity reuse by a suitable partitioning (different from the aforementioned rule) of the set of the pilots.
Third, the coverage probability of a user on a CE pilot decreases with increasing $R_c$ in the higher $\sinr$ regime,  however, the reverse trend is observed for the lower $\sinr$ regime.
\vspace{-0.5cm}
\section{System Model}\label{sec:SysMod}
\vspace{-0.5cm}

\subsection{Network model}
\subsubsection{BS and user locations}
In this work, we analyze the UL performance of a cellular network where each BS is equipped with $M \rightarrow \infty$ antennas. 
The locations of the BSs belong to the set $\Psi_b = \Phi_b \cup \{\nbo\}$, where $\nbo$ represents the origin, and $\Phi_b$ is a realization of homogeneous PPP of density $\lambda_0$.
By virtue of Slivnyak's theorem~\cite{Haenggi2013}, $\Psi_b$ is also a homogeneous PPP of density $\lambda_0$.
The location of the $j$-th BS is denoted by $\bsloc{j} \in \Psi_b$,
where the index $j$ does not represent any ordering and $\bsloc{0} = \nbo = (0, 0)$ is the origin.
In a cell, the region that is within a distance $R_c$ from its BS is defined as the CC region, and hence the CC region for the typical cell at the origin (referred to as $0$-th cell hereafter) is given by
{\small
\begin{align*}
{\cal X}_C(\nbo, R_c, \Psi_b) = \{\nbx \in {\cal V}_{\Psi_b}(\nbo): \|\nbx\| \leq R_c\} = {\cal V}_{\Psi_b}(\nbo) \cap {\cal B}_{R_c}(\nbo), \numberthis
\label{eq:CCReg}
\end{align*}
}%
where 
$
\vCell_{\Psi_b}(\nbo) = \{\nbx \in \nbbR^2: \|\nbx\| \leq \|\nbx - \bsloc{j}\|, \forall \bsloc{j} \in \Psi_b\}
$
is the PVC associated with $\mathbf{b}_0$ and ${\cal B}_{R_c}(\nbo)$ denotes a ball of radius $R_c$ centered at $\nbo$.
Note that the CC regions are equivalent to the JM cells associated with the BSs~\cite{Moller1992}.
These JM cells are usually defined from the perspective of random nucleation and growth process.
However, we follow the definition in \eqref{eq:CCReg} for simpler exposition.
The region of the cell that is beyond $R_c$ from the BS is the CE region and is defined as 
{\small
\begin{align*}
{\cal X}_E(\nbo, R_c, \Psi_b) = \{\nbx \in \vCell_{\Psi_b}(\nbo): \|\nbx\| > R_c\} = \vCell_{\Psi_b}(\nbo) \cap {\cal B}_{R_c}^C(\nbo). \numberthis
\label{eq:CEReg}
\end{align*}
}%
{Note that for $R_c > 0$, there is a non-zero probability that a typical cell may not have a CE region.
We characterize this probability later in Sec.~\ref{sec:AreaDist}.}

The locations of the CC and CE users attached to the $j$-th BS are uniformly and randomly distributed within  ${\cal X}_C(\bsloc{j}, R_c, \Psi_b)$ and ${\cal X}_E(\bsloc{j}, R_c, \Psi_b)$, respectively.
We denote the CC area of the $j$-th cell (or with slight abuse of notation any typical cell) as $\ccAre(\lambda_0, R_c) = |{\cal X}_C(\bsloc{j}, R_c, \Psi_b)|$ and the CE area as $\ceAre(\lambda_0, R_c) = |{\cal X}_E(\bsloc{j}, R_c, \Psi_b)|$.
{If the typical cell does not have a CE region, then ${\cal X}_E(\bsloc{j}, R_c, \Psi_b) = \varnothing$ and $\ceAre(\lambda_0, R_c) = 0$.}
Let $\numUE_{Cj}$ and $\numUE_{Ej}$ be the numbers of CC and CE users present in the $j$-th cell.
We assume that both the random variables $\numUE_{Cj}$  and $\numUE_{Ej}$ follow zero-truncated Poisson distribution with parameters $\lambda_u \ccAre(\lambda_0, R_c)$ and $\lambda_u \ceAre(\lambda_0, R_c)$, respectively.
Hence, conditioned on the CC (CE) area of the $j$-th cell, the probability mass functions of $N_{Cj}$ and $N_{Ej}$ are given as 
\begin{equation}\small{
\dP{N_{Cj} = n| x_c} = \frac{\exp(-\lambda_u x_c) (\lambda_u x_c)^n}{n! (1-\exp(-\lambda_u x_c))}, \normalsize{\text{and}} \
\dP{N_{Ej} = n| x_e, {\cal E}_3^C} = \frac{\exp(-\lambda_u x_e) (\lambda_u x_e)^n}{n! (1-\exp(-\lambda_u x_e))},
\label{eq:UEPMF}}
\end{equation} 
respectively, where ${\cal E}_3^C$ is the event that the $j$-th cell has a CE region and is defined in Section~\ref{sec:AreaDist}.
{The main motivation behind consideration of the truncated Poisson distribution for users is to ensure that each BS in the network has at least one CC and CE user within its Voronoi cell. Since mMIMO will be primarily used for macro cells, from the system perspective, this is a reasonable assumption.
Further, this allows us to model the user point process (to be defined shortly) as a Type-I process introduced in \cite{Haenggi2017} facilitating a rigorous system analysis from the perspective of a typical cell.}
Note that $\lambda_u$ can be used to vary the average load (the number of users per BS) in the network.

Let us define a point process $\Psi_{\tt u, CC}$ that is constructed by randomly and uniformly distributing one point in the CC region of each cell.
Mathematically, this can be expressed as  
{\small
\begin{align*}
\Psi_{\tt u, CC} = \{ U({\cal X}_C(\bsloc{j}, R_c, \Psi_b)): \forall \bsloc{j} \in \Psi_b\},
\end{align*}
}%
where $U(B)$ denotes a uniformly distributed point in $B \subset \R^2$. From the construction, the density of $\Psi_{\tt u, CC}$ is $\lambda_0$.
On the other hand, let $\Psi_{bE}$ denote the set of BSs having a CE region that is defined as 
$\Psi_{bE} = \{\nbb_j: \forall \nbb_j \in \Psi_b, {\cal X}_E(\bsloc{j}, R_c, \Psi_b) \neq \varnothing\}$. 
Now, for the CE case we define the point process $\Psi_{\tt u, CE}$ as 
{\small
\begin{align*}
\Psi_{\tt u, CE} = \{ U({\cal X}_E(\bsloc{j}, R_c, \Psi_b)): \forall \bsloc{j} \in \Psi_{bE}\}.
\end{align*}
}%
Note that since all the BSs in the network may not have a CE region, the density of $\Psi_{\tt u, CE}$ is less than $\lambda_0$. 
Except the users in the typical cell at $\nbo$, rest of the users in the network belong to the interfering cells. 
Let the CC and CE point processes formed by the points in the interfering cells be given as 
{\small
\begin{align*}
\Phi_{\tt u, CC} = \{ U({\cal X}_C(\bsloc{j}, R_c, \Psi_b)) : \forall\bsloc{j} \in \Phi_b\},  \Phi_{\tt u, CE} = \{ U({\cal X}_E(\bsloc{j}, R_c, \Psi_b)) : \forall \bsloc{j} \in \{\Psi_{bE} \setminus \bsloc{0}\}\}.
\end{align*}
}%

\begin{table*}[!h]
\centering
\caption{Summary of Notations}
\scalebox{1}{\small{
\label{tab::notation}
\begin{tabular}{clc}
\hline \hline
\textbf{Notation}                              & \textbf{Description                                                                                                                  } \\
\hline \hline 
$\Psi_b$ and $\lambda_0$                  & Homogeneous PPP modeling the locations BSs and density of $\Psi_b$ \\
$\bsloc{j}$ and $\nbu_{j_k}$							 & Locations of the $j$-th BS and a user attached to the $j$-th BS using $k$-th pilot  \\
$R_c$ and $\kappa = R_c \sqrt{\pi c_2 \lambda_0}$				 & Threshold radius and normalized threshold radius \\
${\cal V}_{\Psi_b}(\bsloc{j})$				 & Voronoi cell associated with the $j$-th BS \\
${\cal X}_C(\bsloc{j}, R_c, \Psi_b)$ and ${\cal X}_E(\bsloc{j}, R_c, \Psi_b)$								 & CC and CE region of the $j$-th cell \\
$\ccAre(\lambda_0, R_c)$ and $\ceAre(\lambda_0, R_c)$								 & CC and CE areas of a typical cell in a network of BS density $\lambda_0$ \\
$\Phi_{\tt u, k}$ and $\ccPhiDen(r, \kappa)$								 & Point processes formed by users using $k$-th pilot and its density function\\
${\cal I}(j,k)$							& Indicator variable that is 1 when $k$-th pilot is used in $j$-th cell \\
${\cal A}_{\tt 0, CC}$ (${\cal A}_{\tt 0, CE}$) 						 & Indicator variable that is 1 when CC (CE) user of interest is assigned a pilot\\
${\cal A}_{\tt 0n, CC}$ (${\cal A}_{\tt 0m, CE}$) 						 & Indicator variable that is 1 when CC (CE) user is assigned the $n$-th ($m$-th) pilot\\
$D_{ij_k}$, and $d_{ij_k}$							& Random distance between the BS at $\bsloc{j}$ and user at $\nbu_{j_k}$, and its realization  \\
$\nbg_{ij_k} \sim {\cal CN}(\mathbf{0}_M, d_{ij_k}^{-\alpha}\nbI_M)$							& Channel vector between $i$-th BS and the user at $\nbu_{j_k}$ \\
$\sinr_{0_k}$			   	& $\sinr$ of the user using the $k$-th pilot in the $0$-th cell \\
$\mathtt{P_{c, k}}$ and $\mathtt{P_{c, l}}$ & Coverage probability of a user using $k$-th CC and $l$-th CE pilot, respectively \\
$B, B_C, B_E$, and $T_c$			   	& Number of total pilots, CC pilots, CE pilots, and coherence time \\
\hline
\end{tabular}
}%
}
\end{table*} 

\subsubsection{Pilot sequences}
We restrict our analysis to a narrowband single-carrier system. Extension to a multi-carrier system is straightforward and hence is skipped in favour of simpler exposition.
In order to successfully decode the data simultaneously transmitted by multiple users in a cell, the BS of the respective cell should possess the CSI of the users in that cell.
In order to get the CSI, in the $j$-th cell, each user is assigned a pilot (sequence) that is selected from a set of orthogonal pilots 
${\cal P}_j \subset {\cal P}$, where ${\cal P} = \{\nbp_1, \nbp_2, \cdots, \nbp_B\}$ and $\nbp_i \in \nbbC^{B \times 1}$ for $i = 1, 2, \ldots, B$, where $B$ is the number of orthogonal pilots.
For simplicity, we denote the pilots by their indices.
Therefore, the set of indicies of the pilots used in the $j$-th cell is denoted as ${\cal K}_j \subset {\cal K}$, where ${\cal K} = \{1, 2, 3, \ldots, B\}$. 
Owing to the limited channel coherence time $T_c$, the cardinality of this set $|{\cal K}| = B \leq T_c$.
While the pilots remain orthogonal in each cell, due to the consideration of FPR, orthogonality among cells is not guaranteed. 
Hence, in each cell, the pilots are partitioned into two different sets, i.e. for the $j$-th BS ${\cal K}_j = \setCCP \cup \setCEP_j$ where $\setCCP$ contains the indices of the CC pilots that are reused in each cell.
Moreover, $|\setCCP| = B_C \leq B$.
On the other hand, $\setCEP_j$ contains the indices of the CE pilots that are reused in other cells in the network depending on the reuse factor.
If $\beta_f$ is the reuse factor of the CE pilot sequence, then $(B - B_C)/\beta_f = B_E = |\setCEP_j|$ for all $\nbb_j \in \Psi_b$. 
The choice for $B_C, B_E$, and $\beta_f$ is made such that all three are integers. 

\begin{figure}[!htb]
\centering
\begin{subfigure}{0.4\linewidth}
  \centering
  \includegraphics[width=1\linewidth]{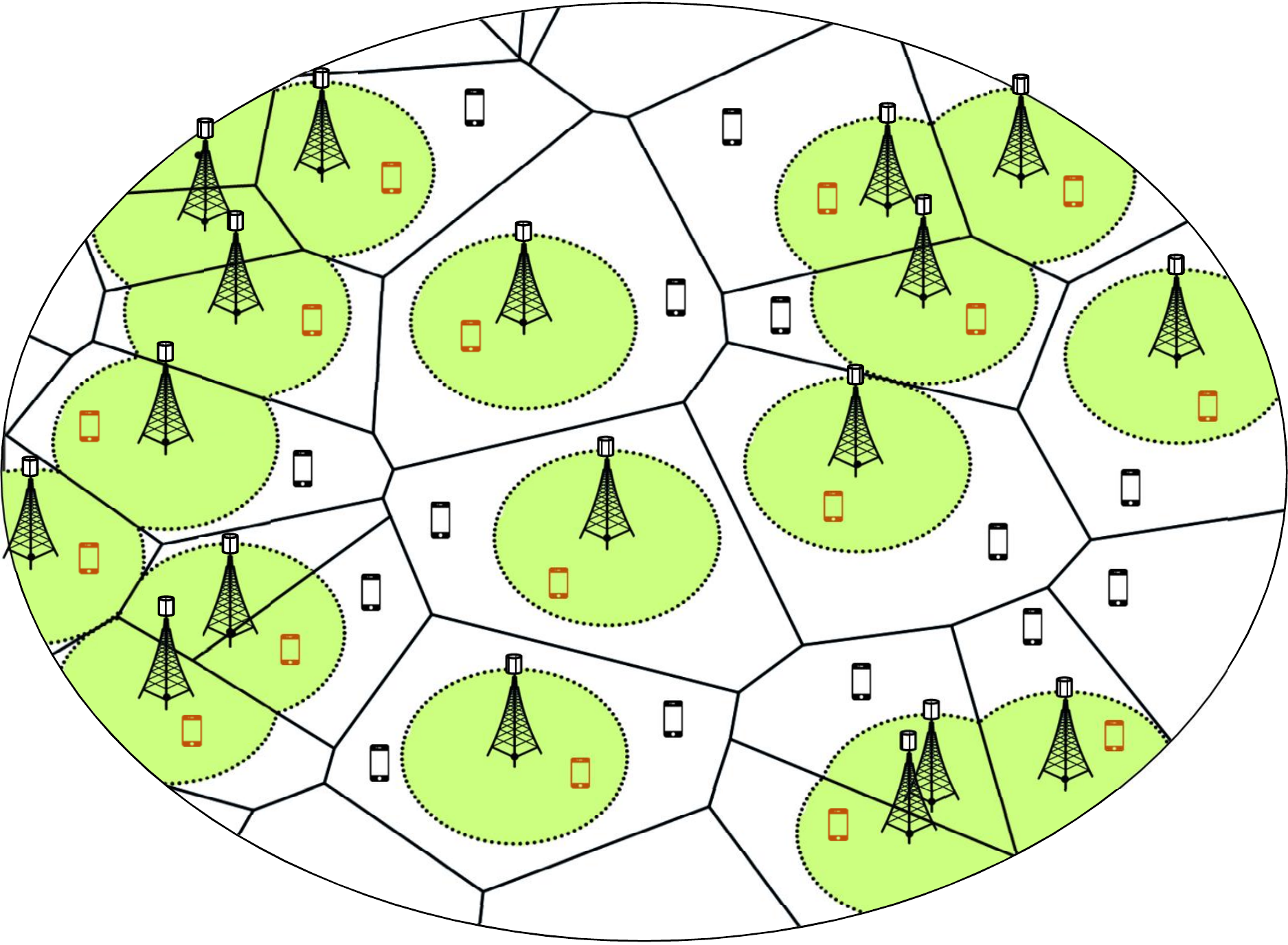}
\end{subfigure}
\hspace{0.75cm}
\begin{subfigure}{0.3\linewidth}
  \centering
  \includegraphics[width=1\linewidth]{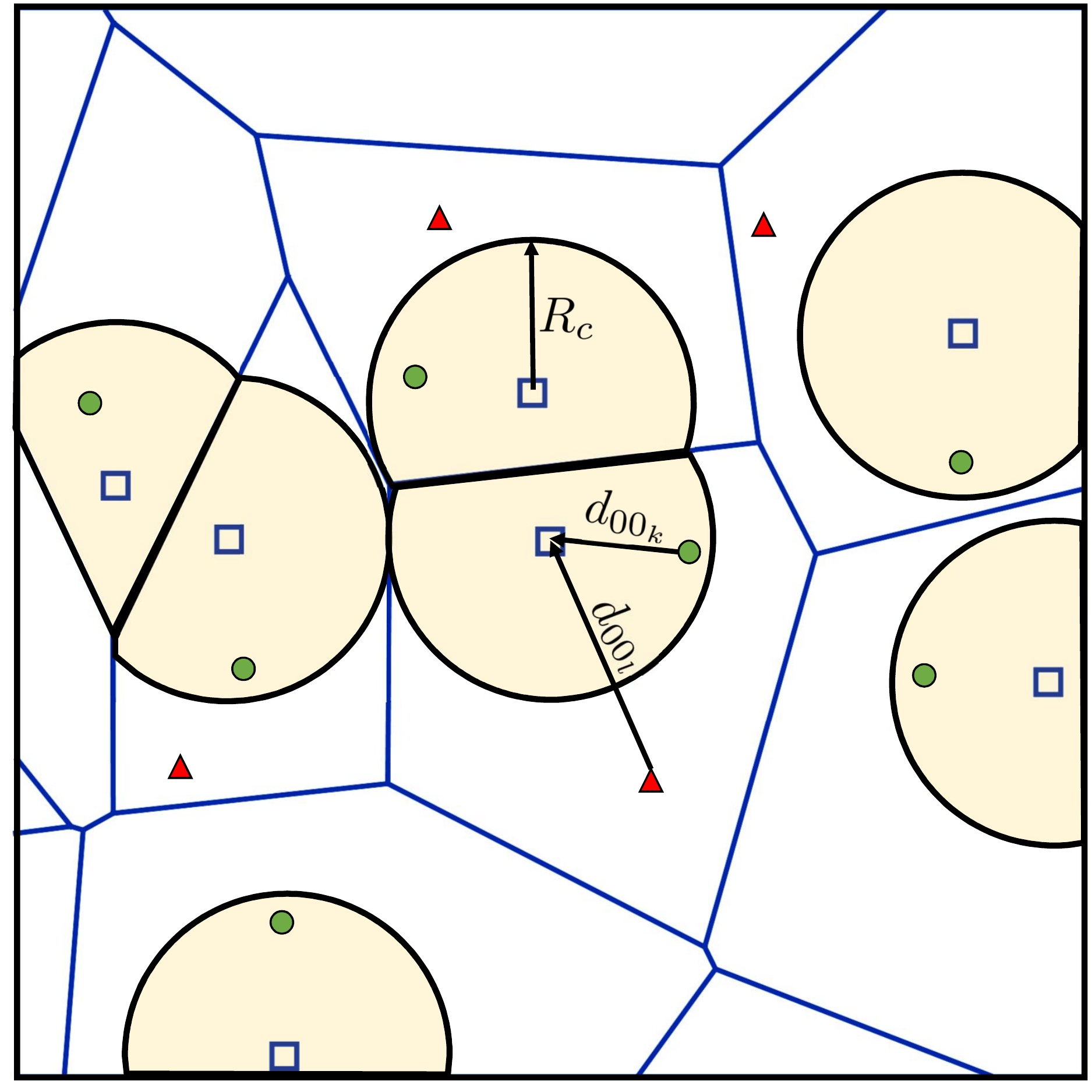}
\end{subfigure}%
\vspace{-0.4cm}
\caption{\footnotesize A representative network diagram (left) and a network realization illustrating the users using the $k$-th CC and $l$-th CE pilot (right). In a few of the cells the CE pilot is not in use.}
\vspace{-0.25cm}
\label{fig:ULDiagram}
\end{figure}

These pilots are assigned randomly to the user in a particular cell. 
Let $k$ be a randomly selected CC pilot. Now, we define a binary random variable ${\cal I}(j,k)$ as follows
\begin{equation}{
{\cal I}(j,k) =
\begin{cases}
0, & \text{if $k$-th pilot sequence is not used in the $j$-th cell}, \\
1, & \text{if $k$-th pilot sequence is used in the $j$-th cell}.
\end{cases}}
\end{equation}
On the similar lines, for a randomly selected CE pilot sequence $l$, we define the binary random variable ${\cal I}(j,l)$.
Let $\Phi_{\tt u, k}$ and $\Phi_{\tt u, l}$ be the point processes formed by the interfering CC and CE users that use the $k$-th CC and $l$-th CE pilots, respectively.
Since the user locations in $\Phi_{\tt u, k}$ are uniformly distributed points in the CC region of their respective cells, $\Phi_{\tt u, k}$ can be defined to inherit the user locations from $\Phi_{\tt u, CC}$ when ${\cal I}(j, k) = 1$. Similar argument is true for $\Phi_{\tt u, l}$ and $\Phi_{\tt u, CE}$.
Hence, we write 
{\small
\begin{align}
\Phi_{\tt u, k} = \{\ueloc{}: \ueloc{} \in \Phi_{\tt u, CC}, {\cal I}(j, k) = 1 \}, \normalsize{\text{and}} \ \Phi_{\tt u, l} = \{\ueloc{}: \ueloc{} \in \Phi_{\tt u, CE}, {\cal I}(j, l) = 1 \}.
\end{align}
}%
We defer the discussion on the statistical properties of these point processes to Section~\ref{sec:PcAndSE}.
Note that the point process formed by the users using other pilot sequences in the network can be defined on the similar lines as that of $\Phi_{\tt u, k} (\Phi_{\tt u, l})$, where the points will be inherited from a point process that has the same definition as $\Phi_{\tt u, CC} (\Phi_{\tt u, CE})$.
{An illustrative network diagram is presented in Fig.~\ref{fig:ULDiagram} with one CC pilot that is reused in each cell and one CE pilot that is reused in a few of the cells.}

\subsubsection{Distance distributions}
Let the location of the user that uses the $k$-th sequence in the $j$-th cell be denoted as $\ueloc{j_k}$.
The random distance between a user at $\ueloc{j_k}$ and a BS at $\bsloc{i}$ is denoted by the random variable $D_{ij_k} = \|\ueloc{j_k} - \bsloc{i}\|$ and $d_{ij_k}$ is its realization.
{In this work, we present the coverage probability of a randomly selected user that is assigned the $k$-th ($l$-th) CC (CE) pilot in the $0$-th cell.
To achieve this goal, the first step is the knowledge of the distribution of serving distance $D_{00_k}$ ($D_{00_l}$) between $\bsloc{0}$ and the CC (CE) user using the $k$-th ($l$-th) pilot.
In case of a typical PVC, the distance distribution between the BS and a randomly located point in the PVC is approximated as Rayleigh distribution with scale parameter $(\sqrt{2 \pi \lambda_0 c_2})^{-1}$, where $c_2 = 5/4$ is an empirically obtained correction factor~\cite{YuMukIshYa2012}.
{Note that the other values of $c_2$ such as 13/10~\cite{Haenggi2017} and 9/7 (the ratio of the mean volumes of  Crofton cell and typical cell) are also good approximations.}
Since, the user at $\nbu_{0_k}$ can not lie beyond ${\cal B}_{R_c}(\nbo)$, it is reasonable to approximate the distribution of $D_{00_k}$ to follow truncated Rayleigh distribution as given below}
{\small
\begin{align*}
F_{D_{00_k}} (d_{00_k}| R_c) = \frac{1 - \exp(- \pi c_2 \lambda_0 d_{00_k}^2)}{1 - \exp(- \pi c_2 \lambda_0 R_c^2)}, \quad d_{00_k} \leq R_c. \numberthis
\label{eq:Djjk}
\end{align*}
}%
On the other hand, the distribution of distance $D_{00_l}$ can also be approximated as
{\small
\begin{align*}
F_{D_{00_l}} (d_{00_l}| R_c) = 1 - \exp(- \pi c_2 \lambda_0 (d_{00_l}^2-R_c^2)), \quad d_{00_l} > R_c. \numberthis
\label{eq:Djjl}
\end{align*}
}%
As it will be evident from the numerical section, these approximated distributions of $D_{00_k}$ and $D_{00_l}$ allow efficient and accurate evaluation of coverage probability. 
At this point, in order to make $R_c$ invariant to the BS density $\lambda_0$, we define a normalized radius $\kappa$ as
$
R_c = \frac{\kappa}{\sqrt{\pi c_2 \lambda_0}}, \kappa \in [0, \infty).
$
In Sec.~\ref{sec:PcAndSE}, $\kappa$ will be used in the statistical characterization of $\Phi_{\tt u,k} (\Phi_{\tt u,l})$. Further, $\kappa$ also provides perspective regarding the size of the CC region without the knowledge of $\lambda_0$.
Next, we define the system parameters from the perspective of the CC user using the $k$-th pilot sequence. 
The extension of these definitions for CE case is straightforward.

\vspace{-0.5cm}
\subsection{Channel model and channel estimation}
\subsubsection{Channel model}
We consider a system where each link suffers from two multiplicative wireless channel impairments, namely distance-dependent pathloss and multi-path fading. 
Consideration of the effect of shadowing is left as a promising future work.
The channel vector between the user located at $\ueloc{j_k}$ and the $M$ antenna elements of the BS located at $\bsloc{i}$ is given as 
$\nbg_{ij_k} = d_{ij_k}^{-\alpha/2} \nbh_{ij_k} (\in \nbbC^{M \times 1}),
$
where $\alpha$ is the pathloss exponent, $\nbh_{ij_k} \sim {\cal CN}(\boldsymbol{0}_M, \nbI_M)$ is a ${M \times 1}$ complex Gaussian vector.
We assume that these channel vectors exhibit quasi-orthogonality, i.e. 
\begin{equation}\small{
\lim_{M \rightarrow \infty} \frac{1}{M} \nbh_{ij_m}^H\nbh_{ij_n} \rightarrow
\begin{cases}
0 & j_m \neq j_n \\
1 & j_m = j_n.
\end{cases}}
\end{equation}
Further, we consider user transmit power $\rho_u$ to be fixed for both pilot and data symbols.

\subsubsection{Channel estimation}
As discussed earlier, orthogonal pilot sequences are assigned to users within a cell for channel estimation.
For simplicity, we assume that each BS obtains the least square channel estimate of the users attached to them. 
Hence, for the CC user using the $k$-th pilot, the channel estimate at the $0$-th BS is given as 
$\tilde{\nbg}_{00_k} = \sqrt{\rho_u} \nbg_{00_k} + \sum_{\ueloc{j_k} \in \Phi_{\tt u,k}} \sqrt{\rho_u} \nbg_{0j_k} + \nbv_0 \in \nbbC^{M \times 1},
$
where $\nbv_0 \sim {\cal CN}(\mathbf{0}_M, \mathbf{I}_M)$ is a complex Gaussian noise vector.

\vspace{-0.5cm}
\subsection{Asymptotic UL $\sinr$ of a CC (CE) user assigned to $k$-th ($l$-th) pilot sequence}
The received signal vector at the $0$-th BS is given as 
\begin{equation}\small{
\nbr_{0} = \nbh_{00_k}x_{0k} d_{00_k}^{-\alpha/2} + \sum_{i=1, i \neq k}^{B} {\cal I}(0, i) \nbh_{00_i} x_{0i} d_{00_i}^{-\alpha/2}  +  \sum_{i=1}^{B} \sum_{\ueloc{j_i} \in \Phi_{\tt u, i}} \nbh_{0j_i}x_{ji} d_{0j_i}^{-\alpha/2} + \nbn_0},
\end{equation}
where $x_{ji}$ is the data symbol transmitted by the user using the $i$-th pilot sequence in the $j$-th cell, $\nbn_0 \sim {\cal CN}(\boldsymbol{0}_M, \nbI_M)$  is a complex Gaussian noise vector.
We assume that $\dE{x_{ji}}{} = 0$ and $\dE{\|x_{ji}\|^2}{}= \rho_u$.
In order to estimate the symbol transmitted by the CC user of interest, the $0$-th BS uses MRC detection scheme, where the filter coefficients are given as $\nbw_{0_k} = \frac{1}{M}\tilde{\nbg}_{00_k}^H$.
As demonstrated in various works in the literature (cf. \cite{Hien2013}), the asymptotic $\sinr$ of a user is independent of the detection scheme used at the BSs.
Now, the detected symbol for the CC user using the $k$-th pilot sequence in the $0$-th BS is given as $\hat{x}_{0k} = \nbw_{0_k} \nbr_0$.
As the number of antennas $M \rightarrow \infty$, due to quasi-orthogonality of the channel, it can be shown that the detected symbol is only affected by the interference from the users using the $k$-th pilot sequence in other cells (a.k.a. pilot contamination). Hence, the $\sinr$ of the CC and CE users that are assigned the $k$-th and $l$-th pilots, respectively, are given as 
\begin{equation}\small{
{\sinr}_{0_k} = d_{00_k}^{-2\alpha}\bigg(\sum_{\ueloc{j_k} \in \Phi_{\tt u, k}} d_{0j_k}^{-2\alpha}\bigg)^{-1}, \normalsize{\text{and}} \ {\sinr}_{0_l} = d_{00_l}^{-2\alpha}\bigg(\sum_{\ueloc{j_l} \in \Phi_{\tt u, l}} d_{0j_l}^{-2\alpha}\bigg)^{-1}.  \numberthis
\label{eq:SinrULCC}}
\end{equation}
The proof of the above $\sinr$ expression is readily available in the literature (cf. \cite{Mar2010, BaiH2014}).
Since the above expressions are independent of $\rho_u$, we assume $\rho_u \equiv 1$.

\vspace{-0.5cm}
\subsection{Performance metrics}
In this work, the following metrics are considered for the network performance analysis.

\subsubsection{$\sinr$ coverage probability} The $\sinr$ coverage probabilities of a CC  and CE user using the $k$-th  and $l$-th pilots for a target $\sinr$ threshold $T$ is defined as
\begin{equation}\small{
{\tt P_{c, k}}(T) = \dP{{\sinr}_{0_k} \geq T | {\cal I}(0, k) = 1}{}, \normalsize{\text{and}} \ {\tt P_{c, l}}(T) = \dP{{\sinr}_{0_l} \geq T | {\cal I}(0, l) = 1, {\cal E}_3^C}{}.}
\end{equation}

\subsubsection{Average user SE} 
{The average user SEs of the CC and CE users of interest are given as 
{\small
\begin{align*}
\overline{\se}_{\tt u, CC} = \left(1 - \frac{B}{T_c}\right) \dE{{\cal A}_{\tt 0, CC}\log_2(1 + {\sinr}_{\tt 0, CC})}{}, \normalsize{\text{and}} \
\overline{\se}_{\tt u, CE} =  \left(1 - \frac{B}{T_c}\right) \dE{{\cal A}_{\tt 0, CE} \log_2(1 + {\sinr}_{\tt 0, CE})\bigg| {\cal E}_3^C}{}, \numberthis
\label{eq:userSE}
\end{align*}
}%
where $(1-B/T_c)$ accounts for the fact that out of the total coherence time of $T_c$ symbol duration, $B$ symbol duration is dedicated for channel estimation leaving only $T_c-B$ duration for data transmission. 
Note that while the coverage probability is defined for a user conditioned on a pilot, the average user SE is defined for a randomly selected CC (CE) user that can be assigned any one of the CC (CE) pilots. Hence, ${\sinr}_{\tt 0, CC}$ and ${\sinr}_{\tt 0, CE}$ is the $\sinr$ of a randomly selected CC (CE) user that we term as {\em CC (CE) user of interest}. 
Further, ${\cal A}_{\tt 0, CC}$ is defined as 
\begin{align*}
{\cal A}_{\tt 0, CC} =
\begin{dcases}
1, & \text{if the CC user of interest is assigned a pilot sequence} \\
0, & \text{otherwise}.
\end{dcases}
\end{align*}
Similarly, we define the indicator variable ${\cal A}_{\tt 0, CE}$ for a random CE user of interest.}

\subsubsection{Average cell SE}
The cell SE of the $0$-th cell is given as 
{\small
\begin{align}
{\cse} = 
\left(1 - \frac{B}{T_c}\right) \left[ \sum_{n=1}^{B_C} \log_2( 1 + \sinr_{0_n}) + \sum_{m=1}^{B_E} \log_2( 1 + \sinr_{0_m})\right], \numberthis
\label{eq:SumCellSE}
\end{align}
}%
where $\sinr_{0_n} = 0 (\sinr_{0_m} = 0)$ if ${\cal I}(0,n) =0 ({\cal I}(0,m) = 0)$.
Our metric of interest is $\dE{{\cse}}{}$.
In the following sections, we derive theoretical expressions for the aforementioned quantities.

\vspace{-0.3cm}
\section{Distributions of the CC and CE areas of a typical cell}\label{sec:AreaDist}
{As discussed in the previous section, the distribution of the number of CC (CE) users and subsequently the pilot utilization in an interfering cell depends on its CC (CE) area. Since exact characterization of CE area is challenging (it is an open problem), we provide an approximate area distribution for the CE area using the well-known Weibull distribution that not only allows faster evaluation but also provides useful insights regarding different performance metrics by leveraging its known statistical properties.
In our approach, we first derive exact expressions for the first two moments of the CE area of a typical cell.
In the second step, using moment matching method, we approximate this area as Weibull distribution.
We use the similar method to approximate the CC area distributions as a truncated beta distribution.
While the exact characterization of the distribution of a typical JM cell area, hence the CC area, is given in \cite{Pineda2007}, the expression of the probability density function ($\pdf$) involves an infinite summation over multi-dimensional integrations. Further, the order of integration (hence the complexity of the expression) increases with the increasing value of $R_c$.
Hence, our approximate truncated beta distribution lends tractability to the analysis.
We validate the accuracy of the proposed distributions through Monte Carlo simulations using statistical metrics such as Kulback-Leibler divergence (KLD) and Kolmogorov-Smirnov  distance (KSD).}

\vspace{-0.5cm}
\subsection{Distribution of CE area of a typical cell}
To begin with, in the following lemma, we present the first two moments of the CE area.
\begin{lemma}\label{lem:M1M2CE}
For a given threshold distance $R_c$ and BS density $\lambda_0$, the mean CE area of a typical Voronoi cell is
{\small
\begin{align*}
m_{1, \ceAre}(\lambda_0, R_c) = \dE{\ceAre(\lambda_0, R_c)}{} = (\exp(- \pi \lambda_0 R_c^2))(\lambda_0)^{-1}, \numberthis
\label{eq:M1CE}
\end{align*} 
}%
and the second moment of the area is
{\small
\begin{align*}
m_{2, \ceAre}(\lambda_0, R_c) = \dE{\ceAre(\lambda_0, R_c)^2}{} = 2\pi \int_{r_1 = R_c}^{\infty} \int_{r_2 = R_c}^{\infty} \int_{u = 0}^{2 \pi} \exp\left(- \lambda_0 V(r_1, r_2, u) \right) {\rm d}u r_2 {\rm d}r_2 r_1 {\rm d}r_1, \numberthis
\label{eq:M2CE}
\end{align*}
}%
where $V(r_1, r_2, u)$ is the area of union of two circles. The radii of these circles are $r_1$ and $r_2$, and the angular separation  between their centers with respect to origin is $u$.
Further,
\begin{equation}\small{
V(r_1, r_2, u) =  r_1^2 \left(\pi - v(r_1, r_2, u) + \frac{\sin(2v(r_1, r_2, u))}{2}\right) +  r_2^2 \left(\pi - w(r_1, r_2, u) + \frac{\sin(2w(r_1, r_2, u))}{2}\right), 
\label{eq:AoUC1C2}}
\end{equation} 
where $v(r_1, r_2, u) = \cos^{-1}\left(\frac{r_1 - r_2 \cos(u)}{\sqrt{r_1^2 + r_2^2 - 2r_1 r_2 \cos(u)}}\right)$, and $w(r_1, r_2, u) = \cos^{-1}\left(\frac{r_2 - r_1 \cos(u)}{\sqrt{r_1^2 + r_2^2 - 2r_1 r_2 \cos(u)}}\right)$.
\end{lemma}
\begin{IEEEproof}
Please refer to Appendix~\ref{app:M1M2CC}.
\end{IEEEproof}

Before proceeding further, some intuition on the type of distribution that provides an accurate approximation is necessary.
Note that a Voronoi cell has two characteristic radii denoted as $R_m$ and $R_M$~\cite{Calka2002}. While $R_m$ corresponds to the radius of the largest circle that completely lies inside a Voronoi cell, $R_M$ is the radius of the smallest circle that encircles a Voronoi cell. 
Using $R_m$ and $R_M$, we define following three disjoint events:
\begin{enumerate}
\item ${\cal E}_1 = \{R_c < R_m \}$, i.e. the CC region completely lies inside the Voronoi cell, 
\item ${\cal E}_2 = \{ R_m \leq R_c < R_M \}$, i.e. the circle ${\cal B}_{R_c}(\nbo)$ and the Voronoi cell $\vCell_{\Psi_b}(\nbo)$ intersects,
\item ${\cal E}_3 = \{R_M \leq R_c \}$,  i.e. there is no CE region.
\end{enumerate} 
In this case, the CE area $\pdf$ can be expressed as the sum of two components:
{\small
\begin{align*}
f_{\ceAre}(x) = f_{\ceAre}(x |{\cal E}_3) \dP{{\cal E}_3}{} + f_{\ceAre}(x |{\cal E}_3^C) (1 - \dP{{\cal E}_3}{}), \numberthis
\label{eq:CEPDF}
\end{align*}
}%
Further, note that $f_{\ceAre}(x |{\cal E}_3)$ is given as 
{\small
\begin{align}
f_{\ceAre}(x |{\cal E}_3) = \delta(0).
\label{eq:PMFCE}
\end{align}
}%
Next we obtain \mbox{$\dP{{\cal E}_3}{}$} and $f_{\ceAre}(x |{\cal E}_3)$. Since \mbox{${\cal E}_3 = \{ R_M \leq R_c \}$}, \mbox{$\dP{{\cal E}_3}{} = \dP{R_M \leq R_c}{}$},
where the distribution of \mbox{$R_M$} is obtained from Theorem~1 of \cite{Calka2002} and is given as 
{\small
\begin{align*}
\dP{R_M \leq r} = 1 - \exp\left(-4 \pi \lambda_0 r^2\right)\left(1 - \sum_{k \geq 1} \frac{(-4\pi\lambda_0 r^2)^k}{k!} \xi_k\right), \quad r > 0. \numberthis
\label{eq:RMCDF}
\end{align*}
}%
In the above expression, 
{\small
\begin{align*}
\xi_k = \int_{\sum\limits_{i=1}^{k} u_i = 1, u_i \in [0, 1]} \left[\prod_{i=1}^k F(u_i) \right] \exp\left(4 \pi \lambda_0 r^2 \sum_{i=1}^k \int_{0}^{u_i} F(t) {\rm d}t\right) {\rm d}\nbu,
\end{align*}
}%
where $F(t) = \sin^2(\pi t) \mathbf{1}(0 \leq t \leq \frac{1}{2}) + \mathbf{1}(t > \frac{1}{2})$, where $\mathbf{1}(\cdot)$ is the indicator function.
Based on moment matching method, we approximate $f_{\ceAre}(x |{\cal E}_3^C)$ as Weibull distribution. Intuitively, the CE area is likely to exhibit similar properties of the Voronoi cell area, especially when $R_c$ is small. Hence, the gamma distribution, which is used to approximate the Voronoi cell area, is the first preference. However, for larger $R_c$, gamma distribution fails to capture the decay of the $\pdf$ of CE area. 
Now, Weibull distribution has similar Kernel as gamma distribution\footnote{The kernel of gamma $\pdf$ is $f_{G}(x) \propto x^{\xi-1} \exp(-x/\theta)$, and Weibull $\pdf$ is $f_{W}(x) \propto x^{\xi-1} \exp(-(x/\theta)^{\xi})$.} along with the flexibility to control the decay factor of the $\pdf$. Therefore, we use Weibull distribution for the aforementioned approximation.
In the following Lemma, we present the mean and variance of $X_E$ conditioned on ${\cal E}_3^C$.
\begin{lemma}\label{lem:CondCEAre}
The mean and variance of the CE area conditioned on ${\cal E}_3^C$ is given as 
{\small
\begin{align*}
 \dE{\ceAre | {\cal E}_3^C}{} =  \dE{\ceAre}{}{}(\dP{{\cal E}_3^C})^{-1}, \quad
 \varn{\ceAre | {\cal E}_3^C} = \varn{\ceAre}  - \dP{{\cal E}_3^C}\dP{{\cal E}_3}(\dE{\ceAre|{\cal E}_3^C}{})^2(\dP{{\cal E}_3^C})^{-1}. \numberthis
 \end{align*}
}%
\end{lemma}
\begin{IEEEproof}
The proof of this Lemma follows from law of total expectation and law of total variance that are given as
$
\dE{\ceAre}{} = \dE{\ceAre | {\cal E}_3}{} \dP{{\cal E}_3}{} + \dE{\ceAre | {\cal E}_3^C}{} \dP{{\cal E}_3^C}, 
$
and
{\small
\begin{align*}
\varn{\ceAre} = & \varn{\ceAre|{\cal E}_3} \dP{{\cal E}_3} + \dP{{\cal E}_3}(1-\dP{{\cal E}_3})(\dE{\ceAre|{\cal E}_3}{})^2
+ \varn{\ceAre|{\cal E}_3^C} \dP{{\cal E}_3^C} \\
& + \dP{{\cal E}_3^C}\dP{{\cal E}_3}(\dE{\ceAre|{\cal E}_3^C}{})^2 - 2 \dE{\ceAre|{\cal E}_3}{} \dP{{\cal E}_3} \dE{\ceAre|{\cal E}_3^C}{}\dP{{\cal E}_3^C}.
\end{align*}
}%
Rearranging the terms, and replacing $\dE{\ceAre|{\cal E}_3}{} = 0$ and $\varn{\ceAre|{\cal E}_3}{} = 0$, we obtain the expressions presented in the Lemma.
\end{IEEEproof}

Note that the $\pdf$ of Weibull distribution is given as 
{\small
\begin{align*}
f_{\ceAre}(x | {\cal E}_3^C) = \frac{\eta}{\zeta} \left(\frac{x}{\zeta}\right)^{\eta-1} \exp\left(-\frac{x^{\eta}}{\zeta^\eta}\right),
\numberthis
\label{eq:WblPDF}
\end{align*} 
}%
where $\eta$ and $\zeta$ are shape and scale parameters. These parameters are obtained by matching the first two moments and solving the following system of equations:
{\small
\begin{align*}
\eta\Gamma(1 + 1/\zeta) =  \dE{\ceAre | {\cal E}_3^C}{}, \quad \eta^2(\Gamma(1 + 2/\zeta) - \Gamma(1 + 1/\zeta)^2) =  \varn{\ceAre|{\cal E}_3^C}. \numberthis
\label{eq:WblParam}
\end{align*}
}%
Replacing the solutions of the above equations in \eqref{eq:WblPDF}, we get the desired $\pdf$ for $f_{X_E}(x|{\cal E}_3^C)$.
Now, \eqref{eq:WblPDF}, \eqref{eq:RMCDF}, \eqref{eq:PMFCE}, and \eqref{eq:CEPDF} together provide us the approximate $\pdf$ for CE area.

\vspace{-0.5cm}
\subsection{Distribution of CC area of a typical cell}
In this section, we present the approximate distribution for the CC area of a typical cell.
Similar to the CE case, in the next lemma, we derive the first two moments of the CC area.
\begin{lemma}\label{lem:M1M2CC}
For a given $\lambda_0$ and $R_c$, the mean CC area of a typical Voronoi cell is given by
\begin{equation}\small{
m_{1, \ccAre}(\lambda_0, R_c) = \dE{\ccAre(\lambda_0, R_c)}{} = (1 - \exp(- \pi \lambda_0 R_c^2))(\lambda_0)^{-1},
\label{eq:M1}}
\end{equation} 
and the second moment of the area is given by 
\begin{equation}\small{
m_{2, \ccAre}(\lambda_0, R_c) = \dE{\ccAre(\lambda_0, R_c)^2}{} = 2\pi \int_{r_1 = 0}^{R_c} \int_{r_2 = 0}^{R_c} \int_{u = 0}^{2 \pi} \exp\left(- \lambda_0 V(r_1, r_2, u) \right) {\rm d}u r_2 {\rm d}r_2 r_1{\rm d}r_1,
\label{eq:M2}}
\end{equation}
where \mbox{$V(r_1, r_2, u)$} is the area of union of two circles given in \eqref{eq:AoUC1C2}.
\end{lemma}
\begin{IEEEproof}
The proof can be done on the similar lines as that of Lemma~\ref{lem:M1M2CE}.
\end{IEEEproof}


Similar to the CE case, the $\pdf$ of the CC area can be expressed as 
{\small
\begin{equation}
f_{\ccAre}(x) = f_{\ccAre}(x |{\cal E}_1) \dP{{\cal E}_1}{} + f_{\ccAre}(x |{\cal E}_1^C) (1 - \dP{{\cal E}_1}{}),
\label{eq:pdfXc1}
\end{equation}
}%
where $\dP{{\cal E}_1}{} = \dP{R_m > R_C}{}$. Note that $R_m$ is half of the nearest neighbour distance of a PPP that follows a Rayleigh distribution with parameter $(\sqrt{8 \pi \lambda_0})^{-1}$ and is given as
\begin{equation}\small{
F_{R_m}(r_m) = 1 - \exp(-4 \pi \lambda_0 r_m^2).}
\label{eq:CDFRm}
\end{equation}
Now, the probability of ${\cal E}_1$ is given as 
\begin{equation}\small{
\dP{{\cal E}_1}{} = \dP{R_m > R_c} = \exp(-4 \pi \lambda_0 R_c^2) =  1 - \dP{{\cal E}_1^C}{}. \numberthis
\label{eq:PE1}}
\end{equation}
Observe that, the $\pdf$ of $\ccAre$ conditioned on ${\cal E}_1$ is
\begin{equation}\small{
f_{\ccAre}(x |{\cal E}_1) = \delta(\pi R_c^2),
\label{eq:pdfAcE1}}
\end{equation}
where $\delta(x)$ is the Dirac-delta function.
The final part that remains to be determined in the above equation is a suitable approximate distribution for $f_{\ccAre}(x |{\cal E}_1^C)$. 
In this work, we approximate $f_{\ccAre}(x |{\cal E}_1^C)$ by generalized truncated beta distribution which is given as 
{\small
\begin{align*}
& f_{\ccAre}(x |{\cal E}_1^C) \approx  g(x; v, w, y, z, \alpha, \beta) 
= \frac{(x-y)^{\alpha-1} (z-x)^{\beta-1}}{B(v, w, y, z; \alpha, \beta)}, \quad 0 \leq x  < \pi R_c^2, \numberthis
\label{eq:GenTrBeta}
\end{align*}
}%
where $\alpha$ and $\beta$ are shape parameters; the support of the untruncted beta distribution is \mbox{\small $[y, z]$} (since beta distribution has finite support); the support of the truncated beta distribution is \mbox{\small$[v, w]$}; and the normalization factor \mbox{\small $B(v, w, y, z; \alpha, \beta)$} is {\small
\begin{align*}
B(v, w, y, z; \alpha, \beta) = \int_{\mathfrak{v}}^{\mathfrak{w}} (x-y)^{\alpha-1} (z-x)^{\beta -1} {\rm d}x,
\end{align*}
}%
where \mbox{$\mathfrak{v} = \frac{v-y}{y-z}$} and \mbox{ $\mathfrak{w} = \frac{w-y}{z-y}$}.
The choice of beta distribution is motivated by the fact that the distribution  function of \mbox{ $X_c$} has a finite support \mbox{ $[0, \pi R_c^2]$}.
Based on this support set, we set \mbox{$v = 0$ and $w = \pi R_c^2$} for the $\pdf$ presented in \eqref{eq:GenTrBeta}.
Another motivation behind selection of beta is the presence of an additional shape parameter compared to conventional distributions such as Gamma or Weibull, which are parametrized by a single shape parameter.
Further, we are introducing truncation to the above distribution that gives us an additional degree of freedom to closely match any arbitrary shape of the actual $\pdf$.
{Here, we set \mbox{$y = 0$} and \mbox{$z = 3/2 \pi R_c^2$}.}
To obtain the shape parameters $\alpha$ and $\beta$, we follow the moment matching method, for which we need the mean and variance of $\ccAre$ conditioned on ${\cal E}_1^C$. In the following Lemma, we derive aforementioned mean and variance.
\begin{lemma}\label{lem:CondCCAre}
The mean and variance of the area $\ccAre$ conditioned on ${\cal E}_1^C$ is given as
\begin{equation}\small{
\begin{aligned}
 \dE{\ccAre | {\cal E}_1^C}{} = &  \left(\dE{\ccAre}{} - \dE{\ccAre | {\cal E}_1}{} \dP{{\cal E}_1}{}\right)/(1-\dP{{\cal E}_1}), \\
 \varn{\ccAre | {\cal E}_1^C} = & \frac{1}{(1-\dP{{\cal E}_1})} \big(\varn{\ccAre} - \dP{{\cal E}_1}(1-\dP{{\cal E}_1})(\dE{\ccAre|{\cal E}_1}{})^2 \\ & - \dP{{\cal E}_1^C}\dP{{\cal E}_1}(\dE{\ccAre|{\cal E}_1^C}{})^2 + 2 \dE{\ccAre|{\cal E}_1}{} \dP{{\cal E}_1} \dE{\ccAre|{\cal E}_1^C}{}\dP{{\cal E}_1^C} \big).
\end{aligned}}
\end{equation}
\end{lemma}
\begin{IEEEproof}
The proof of this Lemma can be done on the similar lines as that of Lemma~\ref{lem:CondCEAre} and using the fact $\varn{\ccAre|{\cal E}_1} = 0, \dE{\ccAre | {\cal E}_1}{}  = \pi R_c^2$.
\end{IEEEproof}

The parameters $\alpha, \beta$ in \eqref{eq:GenTrBeta} are obtained by solving the following simultaneous equations
{\small
\begin{align*}
\frac{\mathtt{B}(v, w, y, z; \alpha+1, \beta)}{\mathtt{B}(v, w, y, z; \alpha, \beta)} =  \dE{\ccAre | {\cal E}_1^C}{}, \quad \frac{\mathtt{B}(v, w, y, z; \alpha+2, \beta)}{\mathtt{B}(v, w, y, z; \alpha, \beta)} - \dE{\ccAre | {\cal E}_1^C}{}^2  =  \varn{\ccAre|{\cal E}_1^C}. \numberthis
\label{eq:CCSysEq}
\end{align*}
}
Substituting \eqref{eq:PE1} and \eqref{eq:pdfAcE1} in \eqref{eq:pdfXc1}, the  approximate CC area $\pdf$ is given as  
{\small
\begin{align*}
f_{\ccAre}(x) = \delta(\pi R_c^2) \exp(-4 \pi \lambda_0 R_c^2) + f_{\ccAre}(x |{\cal E}_1^C) (1 -  \exp(-4 \pi \lambda_0 R_c^2)), \numberthis
\label{eq:PDFCCArea}
\end{align*}
}%
where $f_{\ccAre}(x |{\cal E}_1^C)$ is given in \eqref{eq:GenTrBeta}.
\vspace{-0.5cm}
\subsection{Accuracy of the approximate distributions}
\begin{figure}[!htb]
\centering
\begin{subfigure}{0.4\textwidth}
  \centering
  \includegraphics[width=1\linewidth]{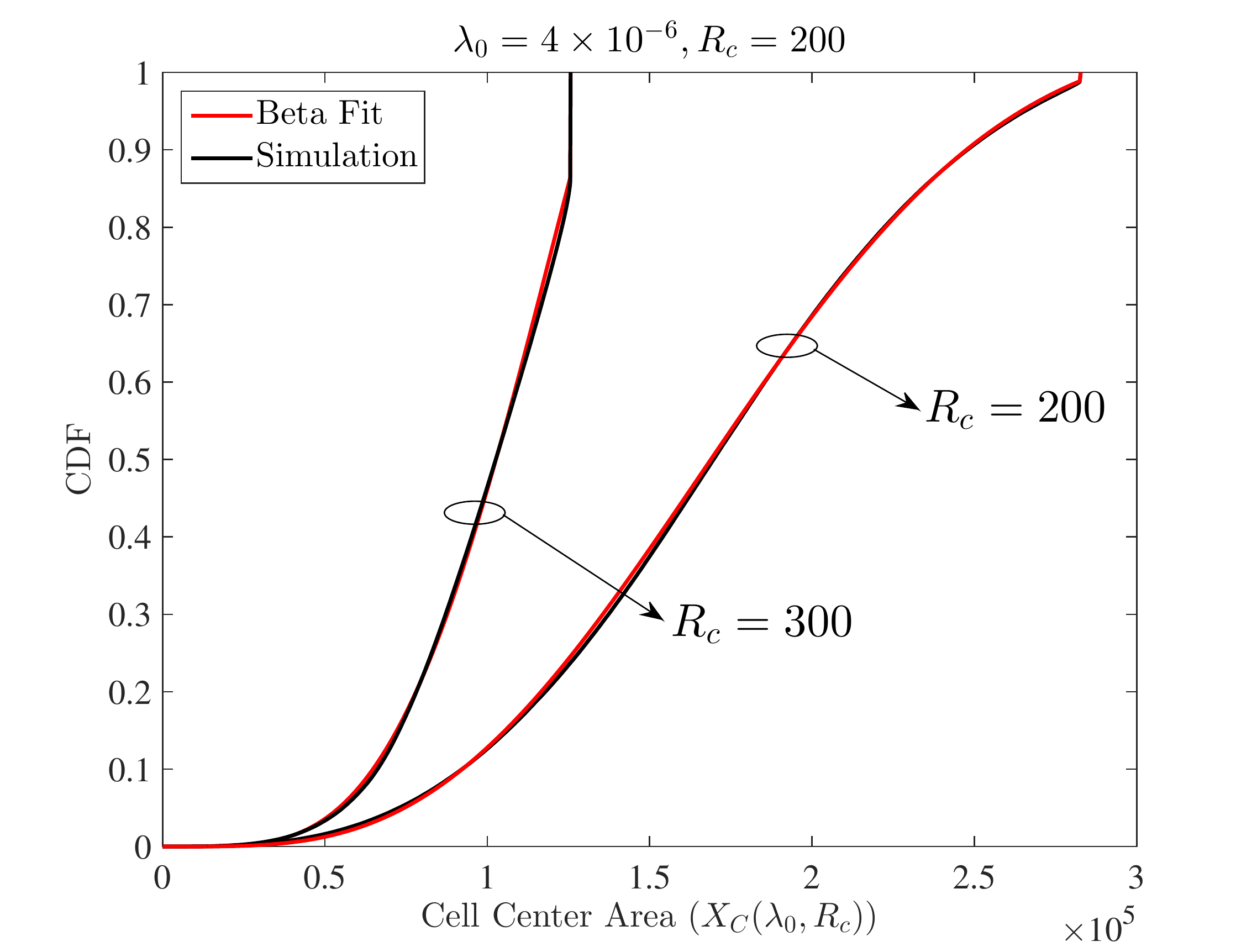}
\end{subfigure}
\begin{subfigure}{0.4\textwidth}
  \centering
  \includegraphics[width=1\linewidth]{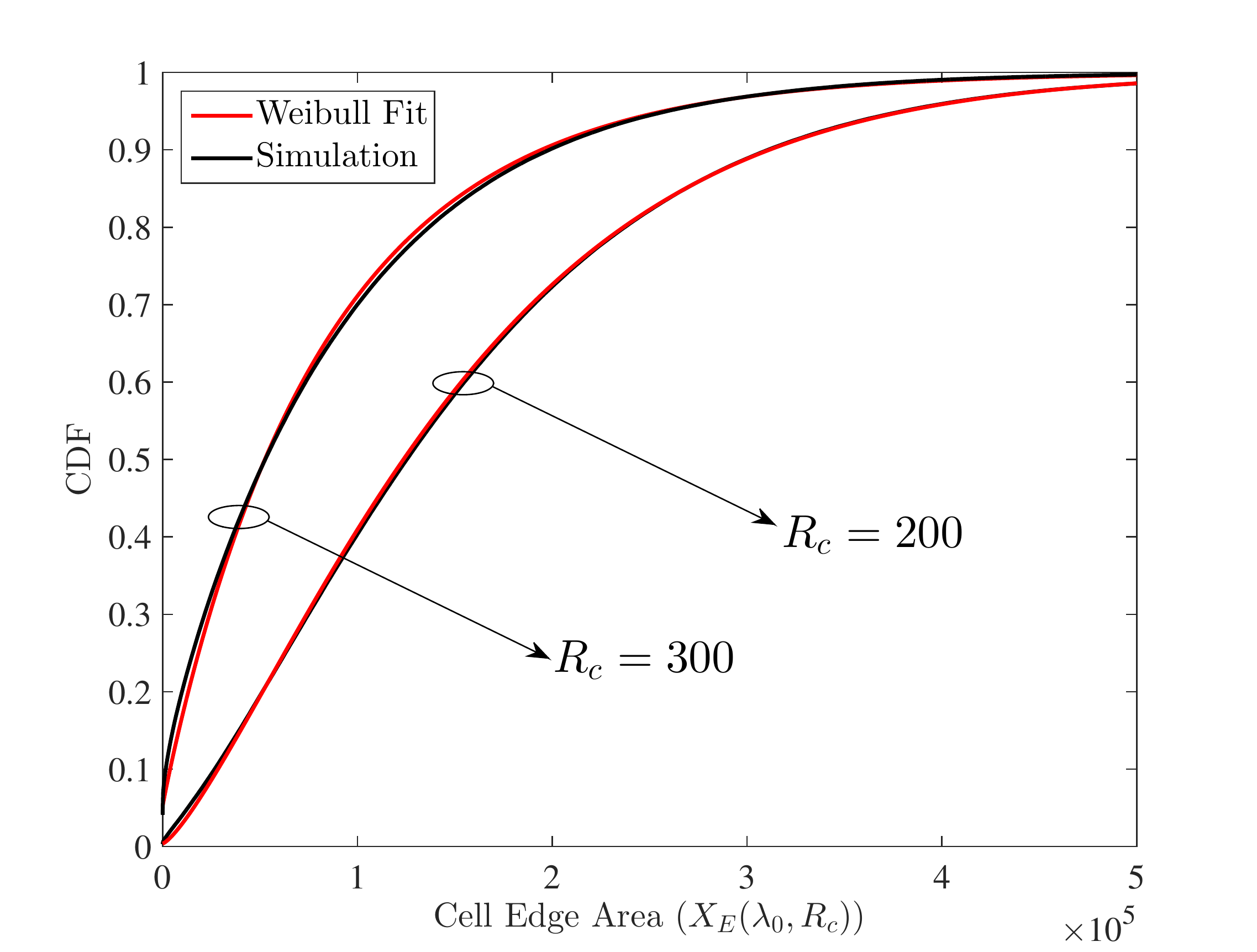}
\end{subfigure}%
\vspace{-0.25cm}
\caption{\small (Left) The $\cdf$ of CC area. (Right) The $\cdf$ of CE area. $\lambda_0 = 4 \times 10^{-6}$.}
\vspace{-0.25cm}
\label{fig:SimThCom}
\end{figure}

The approximate theoretical results are validated through Monte Carlo simulations.
We use KL divergence (and KS distance) to compare the approximate and the true $\pdf$s ($\cdf$s) obtained through simulations. 
In Table~\ref{tab:KSKL} these two metrics are presented for different values of $R_c$ for both CC and CE areas.
As observed from the table, KSD and KLD are low for all values of $R_c$ verifying the accuracy of the distributions.
In Fig.~\ref{fig:SimThCom}, we compare the true and approximate $\cdf$s of the CC and CE area for visual verification purpose. 

\begin{table}[!htb]
\centering{\small
\begin{tabular}{|c|l|l|l|c|l|}
\hline
\multicolumn{1}{|c|}{$R_c | \kappa$}                           & \multicolumn{1}{c|}{$100 | 0.4$} & \multicolumn{1}{c|}{$200 | 0.8$} & \multicolumn{1}{c|}{$250 | 1$} & $300 | 1.2$     & $500 | 2$    \\ \hline
KS Distance (CC)  & 0.0230                   & 0.0238                   & 0.0123                   & \multicolumn{1}{l|}{0.0104} & 0.002  \\ \hline
KL Divergence (CC) & 0.0125                   & 0.0095                   & 0.0055                   & 0.0032                      & 0.0007 \\ \hline
KS Distance (CE)  & 0.0164                   & 0.0107                   & 0.0233                   & 0.0347 & \\ \hline
KL Divergence (CE) & 0.0098                   & 0.0087                   & 0.0160                   & 0.0208  &                    \\ \hline
\end{tabular}}
\vspace{-0.25cm}
\caption{\small Comparison between simulation and approximate $\pdf$s and $\cdf$s for different $R_c$. $\lambda_0 = 4 \times 10^{-6}$.}
\vspace{-0.25cm}
\label{tab:KSKL}
\end{table}

\vspace{-0.5cm}
\section{Pilot Assignment and Pilot Utilization Probability}\label{sec:PilotUtil}
Leveraging the area distribution presented in the previous section, in this section, we present theoretical expressions for (a) the probability of assigning the $k$-th ($l$-th) pilot sequence to a randomly selected CC (CE) user of interest (Lemma~\ref{lem:PilotAssgnCC}),  and (b) the probability that the $k$-th ($l$-th) pilot sequence is being used in the $j$-th cell (Lemma~\ref{lem:PilotUsageCC}). 
As we will see in the following section, the former quantity is useful in obtaining the average SE of a randomly selected CC (CE) user, and the latter quantity is useful in determining the average cell SE as well as the density function of interfering CC (CE) user point process.
Before proceeding further, let us define ${\cal A}_{\tt 0n, CC}$ as 
{ 
\begin{align*}
{\cal A}_{\tt 0n, CC} =
\begin{dcases}
1, & \text{if the CC user of interest is assigned the $n$-th pilot sequence} \\
0, & \text{otherwise}.
\end{dcases}
\end{align*}
}%
Similarly, the indicator variable ${\cal A}_{\tt 0m, CE}$ can be defined for CE user of interest and the $m$-th CE pilot. Next, we present the probability of pilot assignment to the CC (CE) user of interest. 
\begin{lemma}\label{lem:PilotAssgnCC}
The probability that a randomly selected CC user is assigned the $k$-th pilot is
{\small
\begin{align*}
& \dE{{\cal A}_{\tt 0k, CC}}{}  = \dP{{\cal A}_{\tt 0k, CC} =  1}{} = B_C^{-1} \dP{{\cal A}_{\tt 0, CC} =  1}{} = B_C^{-1}\int\limits_0^{\pi R_c^2} \dP{{\cal A}_{\tt 0, CC} = 1 |x_c} f_{X_C}(x_c) {\rm d}x_c, \quad \normalsize{\text{where}} \\
& \dP{{\cal A}_{\tt 0, CC} =1| x_c}  =  \sum_{n=1}^{B_C} \dP{N_{C0} = n | x_c}   + \sum_{n > B_C} \frac{B_C}{n} \dP{N_{C0} = n | x_c}\numberthis
\label{eq:PilotAssgnProb}
\end{align*}
}%
is the probability that CC user of interest is assigned a pilot in the $0$-th cell.
Further, conditioned on the event that the $0$-th cell has a CE region,
the probability of that a randomly selected CE user is assigned the $l$-th pilot is given as
{\small
\begin{align}
\dE{{\cal A}_{\tt 0l, CE}|{\cal E}_3^C}{} = \dP{{\cal A}_{\tt 0l, CE} = 1| {\cal E}_3^C}{} = \frac{\dP{{\cal A}_{\tt 0, CE} = 1| {\cal E}_3^C}{}}{B_E} = \frac{\int\limits_0^{\infty} \dP{{\cal A}_{\tt 0, CE} = 1 | {\cal E}_3^C, x_e} f_{X_E}(x_e|{\cal E}_3^C) {\rm d}x_e}{B_E},
\end{align}
}%
where {\small
$
\dP{{\cal A}_{\tt 0, CE} =1 | {\cal E}_3^C, x_e} = \sum_{n=1}^{B_E} \dP{N_{E0} = n| {\cal E}_3^C, x_e}   + \sum_{n > B_E} \frac{B_E}{n} \dP{N_{E0} = n |{\cal E}_3^C, x_e}.
$}
\end{lemma}

\begin{IEEEproof}
The probability of assigning a pilot to the CC user of interest is given as 
{\small
\begin{align*}
\dP{{\cal A}_{\tt 0, CC}=1} = \dP{\cup_{n=1}^{B_C}\{{\cal A}_{\tt 0n, CC}=1\}} = \sum_{n=1}^{B_C} \dP{{\cal A}_{\tt 0n, CC}=1} = B_C \dP{{\cal A}_{\tt 0k, CC}=1},
\end{align*}
}%
where the last step follows from the fact that the events $\{\{{\cal A}_{\tt 0n, CC}=1\}, n=1, \ldots, B_C\}$ are equi-probable.
Conditioned on the CC area of the $0$-th cell, the distribution of the number of users in this region is given by \eqref{eq:UEPMF}.
Hence, the probability that the CC user of interest is assigned a pilot is given by~\eqref{eq:PilotAssgnProb}.
The final result is obtained by de-conditioning w.r.t. CC area of the $0$-th cell.
The pilot assignment probability for the CE user follows from the similar argument.
\end{IEEEproof}

As discussed in Sec.~\ref{sec:SysMod}, since our analysis is performed for the $k$-th ($l$-th) pilot, the aggregate network interference perceived at the $0$-th BS depends on the utilization of the $k$-th ($l$-th) pilot in the interfering cells. 
In the following Lemma, we present the probability of the usage of the $k$-th ($l$-th) pilot in an interfering cell.

\begin{lemma}\label{lem:PilotUsageCC}
The probability that the $k$-th pilot is used in an interfering cell (say $j$-th cell) is
{\small
\begin{align}
& \dE{{\cal I}(j, k)}{} = \dP{{\cal I}(j, k) = 1}{} = \int_{0}^{\pi R_c^2} \dP{{\cal I}(j, k) = 1| x_c}{} f_{\ccAre}(x_c) {\rm d}x_c, \ \normalsize{\text{where}}
\label{eq:EIjk}
\end{align}
}%
{\small
\begin{align}
\dP{{\cal I}(j, k)  = 1 | x_c} =  \sum_{n=1}^{B_C} \frac{n}{B_C} \dP{N_{Cj} = n | x_c}   + \sum_{n > B_C} \dP{N_{Cj} = n | x_c}.
\label{eq:TempL5}
\end{align}
}%
Similarly, conditioned on the event that the $j$-th cell has a CE region, the probability that the $l$-th CE pilot is used in the $j$-th cell is given as 
{\small
\begin{align}
\dE{{\cal I}(j, l)|{\cal E}_3^C}{} = \dP{{\cal I}(j, l) = 1|{\cal E}_3^C}{} = \int_{x_e=0}^{\infty} \dP{{\cal I}(j, l) = 1|{\cal E}_3^C, x_e}{} f_{\ceAre}(x_e| {\cal E}_3^C) {\rm d}x_e, \normalsize{\text{where}}
\label{eq:EIjl}
\end{align}
}%
{\small
\begin{align}
\dP{{\cal I}(j, l) = 1|{\cal E}_3^C, x_e}{} = \sum_{n=1}^{B_E} \frac{n}{B_E} \dP{N_{Ej} = n | x_e, {\cal E}_3^C}   + \sum_{n > B_E} \dP{N_{Ej} = n| x_e, {\cal E}_3^C}.
\end{align}
}%
\end{lemma} 
\begin{IEEEproof}
For the CC case, first we condition on area of the $j$-th cell. Now, the probability that the $k$-th sequence is used on the $j$-th cell is given by \eqref{eq:TempL5}. The expression in \eqref{eq:EIjk} follows from de-conditioning w.r.t. $X_c$. On the similar lines, \eqref{eq:EIjl} can be derived.
\end{IEEEproof}


\vspace{-0.5cm}
\section{$\sinr$ Coverage and SE Analysis}\label{sec:PcAndSE}
In Sec.~\ref{sec:SysMod}, we introduced the point processes $\Phi_{\tt u, k}$ and $\Phi_{\tt u, l}$, which are essential for the coverage analysis.
In this section, we characterize the statistical properties of these point processes and subsequently use them to obtain the statistics of network interference observed at the $0$-th BS. 

\vspace{-0.25cm}
\subsection{$\sinr$ coverage analysis of a user assigned to the $k$-th CC pilot}
As discussed in Sec.~\ref{sec:SysMod}, $\Phi_{\tt u, k}$ is obtained from $\Phi_{\tt u, CC}$.
Therefore, the first step is to understand the properties to $\Phi_{\tt u, CC}$ that is discussed next.

\subsubsection{Density function of $\Phi_{\tt u, CC}$}
Conditioned on the $0$-th BS location, $\Phi_{\tt u, CC}$ is isotropic. In addition, since $\Phi_{\tt u, CC}$ is defined excluding the point in ${\cal X}_c(\nbo, R_c, \Psi_b)$ from $\Psi_{\tt u, CC}$, it is non-homogeneous.
Now, our objective is to characterize $\Phi_{\tt u, CC}$ conditioned on the $0$-th BS location $\nbo$.
To achieve this objective, we first determine the PCF $g(r)$ of the points in $\Phi_{\tt u, CC}$ w.r.t. $\nbo$.
Next, using this PCF, we approximate the point process as a non-homogeneous PPP. 
{Although the approach that we have followed for the statistical characterization of $\Phi_{\tt u, CC}$ is inspred by the work presented in~\cite{Haenggi2017}, our result is slightly more general, i.e. for a sufficiently large value of $\kappa$ we arrive at the result presented in~\cite{Haenggi2017}.
Further, as we will see shortly, the derivation of the PCF is also not straightforward as the geometry of the region that we encounter is a little more complex compared to the Voronoi cells considered in~\cite{Haenggi2017}.}
Note that in this case, the PCF $g_{\lambda}(r, \kappa)$ is also a function of $\kappa$. By definition, $g_{\lambda}(r, \kappa)$ presents the likelihood of finding a point of $\Phi_{\tt u, CC}$ at a distance $r$ from the $0$-th BS in a network with $\lambda_0 = \lambda$ and threshold radius $R_c = \kappa/\sqrt{\pi c_2 \lambda}$. 
Further, in this case, the PCF is scale-invariant, i.e. $g_{\lambda}(r, \kappa) = g_1(r \sqrt{\lambda}, \kappa)$.
Using the scale invariance property, in the following Lemma, we present the PCF of $\Phi_{\tt u, CC}$ w.r.t. origin for $\lambda_0 = 1$. 
\begin{lemma}\label{lem:PCFCC}
The PCF of $\Phi_{\tt u, CC}$ w.r.t. the $0$-th BS location is given as 
{\small
\begin{align*}
g_1^{\rm CC}(r, \kappa) \approx 1 - \exp\left(-2 \pi r^2 \dE{\ccAre(1, \kappa/\sqrt{\pi c_2})^{-1}}{}\right), \numberthis
\label{eq:PCFCC}
\end{align*}
}%
where \mbox{\small $\ccAre(1, \kappa/\sqrt{\pi c_2})$} is the CC area of a typical cell of a PV tessellation with unity BS density. 
\end{lemma}
\begin{IEEEproof}
Please refer to Appendix~\ref{app:PCFCC}.
\end{IEEEproof}

Using the above PCF, we approximate $\Phi_{\tt u, CC}$ as a non-homogeneous PPP  such that for all $f : \R^2 \mapsto \R^+$
{\small
\begin{align*}
\mathbb {E}\bigg[ \sum _{x\in \Phi_{\tt u, CC}} f(x) \bigg] \equiv \mathbb {E} \bigg[\sum _{x\in \Phi_{\tt u, CC}^{\mathrm {PPP}} } f(x)\bigg] 
\implies  \lambda_0 \int_{\nbx \in \R^2} f(\nbx) g_1^{\rm CC}(\|\nbx\| \sqrt{\lambda_0}, \kappa) {\rm d}\nbx =  \int_{\nbx \in \R^2} f(\nbx) \lambda_{\tt u, CC}^{\mathrm{PPP}}(\|\nbx\|, \kappa) {\rm d}\nbx,
\end{align*} 
}%
where the second step follows from the application of Campbell's theorem and replacing the intensity measure by the reduced second factorial moment measure~\cite[Chapter~8]{Haenggi2013}. Hence, the density function of $\Phi_{\tt u, CC}$, if approximated as a non-homogeneous PPP, is given as 
{\small
\begin{align*}
\lambda_{\tt u, CC}^{\mathrm{PPP}}(r, \kappa) =
 \lambda_0\left(1 -\exp\big(- 2 \pi \lambda_0 r^2 \dE{X_C(1, \kappa/\sqrt{\pi c_2})^{-1}}{}\big)\right). \numberthis
\label{eq:IntUEDen}
\end{align*}
}%
\subsubsection{Density function of $\Phi_{\tt u,k}$}
Since \mbox{$\Phi_{\tt u, k} \subseteq \Phi_{\tt u, CC}$}, one can obtain $\Phi_{\tt u, k}$ by independently thinning the points in \mbox{$\Phi_{\tt u, CC}$} with probability \mbox{$1-\dE{{\cal I}(k, j)}{}$.} 
Note that due to correlation in CC areas of neighbouring cells, the number of users in each cell, as well as the pilot utilization probability among neighbouring cells are correlated.
Hence, the independent thinning used above is an approximation. However, to maintain tractability, this approximation is necessary.
Approximating $\Phi_{\tt u, k}$ as a PPP $\ccPhi$, in the following Lemma, we present its density function.
\begin{lemma}\label{lem:DenFunCC}
The density function of \mbox{\small $\ccPhi$} is given as 
{\small
\begin{align*}
\ccPhiDen(r, \kappa) =  \lambda_0 \dE{{\cal I}(k,j)}{} \left(1 - \exp\left(- 2 \pi \lambda_0 r^2 \dE{X_C(1, \kappa/\sqrt{\pi c_2})^{-1}}{}\right)\right), \numberthis
\end{align*}
}%
where \mbox{\small $\dE{{\cal I}(k, j)}{}$} is presented in Lemma~\ref{lem:PilotUsageCC}. Corresponding intensity measure is given as 
{\small
\begin{align*}
\ccPhiIM(r, \kappa) = 2 \pi \int_{0}^{r} \ccPhiDen(t, \kappa) t {\rm d}t.
\end{align*}
}%
\end{lemma}
\begin{IEEEproof}
By independently thinning \mbox{\small $\Phi_{\tt u, CC}^{\mathrm{PPP}}$} with probability $1-\dE{{\cal I}(k, j)}{}$, we arrive at the expression for the density function.
\end{IEEEproof}

\subsubsection{Coverage probability of the CC user of interest}
In stochastic geometry-based works, for coverage analysis, one key intermediate step is to characterize the interference by the Laplace transform ($\lt$) of its distribution~\cite{AndGupJ2016}. 
The main advantage of this approach is that in the presence of exponential fading gain, the coverage probability can be readily expressed in terms of this $\lt$ expression~\cite{AndGupJ2016}. 
However, in the $\sinr$ expression given in \eqref{eq:SinrULCC}, the small scale fading term is absent due to spatial averaging.
Hence, the conventional $\lt$ based approach is not applicable in this scenario.
Although classical approaches such as Gil-Palaez inversion theorem  \cite{Gil1951,RenGua2014} can be used to obtain coverage probability, it is computationally inefficient, hence, usually avoided wherever possible.
A more useful solution to this problem can be obtained by observing the fact that due to pathloss the total interference is likely to be dominated by interference contributions from a few {\em dominant} users~\cite{SchloemannDB2016}.
Based on this intuition, we approximate the total interference power as the sum of the interference power from the {\em most} dominant interferer and the mean of the rest of the terms conditioned on the dominant term.

{\em Dominant interferer approximation:}
Since the intensity measure \mbox{$\ccPhiIM(r, \kappa)$} of the point process \mbox{$\ccPhi$} is available, the $\cdf$ of the distance to the nearest interferer is obtained from the void probability of the PPP~\cite{AndGupJ2016}. 
Let \mbox{$\hat{D}_{01_k}$}  be the distance between the $0$-th BS and its nearest interferer. 
Then, the $\cdf$ and the $\pdf$ of \mbox{$\hat{D}_{01_k}$} are given as 
{\small
\begin{align}
F_{\hat{D}_{01_k}}(\hat{d}_{01_k}|\kappa) = 1 - \exp(-\ccPhiIM(\hat{d}_{01_k}, \kappa)), \quad
f_{\hat{D}_{01_k}}(\hat{d}_{01_k}|\kappa) & = 2\pi \hat{d}_{01_k} \ccPhiDen(\hat{d}_{01_k}, \kappa)\exp(-\ccPhiIM(\hat{d}_{01_k},\kappa)).
\label{eq:PDFCDFDomIntDisThrm}
\end{align}
}%
As discussed earlier, the aggregate interference can be approximated as the sum of interference from the most dominant interferer and the expected interference from rest of the interferers in the network.
Hence, we write 
{\small
\begin{align*}
I_{{\tt agg}, k} = \hat{D}_{01_k}^{-2\alpha} + \nbbE\bigg[\sum_{{\hat{\bf u}_{j_k}} \in \ccPhi \setminus {\hat{\bf u}_{1_k}}} \hat{D}_{0j_k}^{-2\alpha} \bigg| \hat{D}_{01_k}\bigg] = \hat{D}_{01_k}^{-2\alpha} + \dE{I_{\tt rem, k}| \hat{D}_{01_k}}{},\numberthis
\label{eq:AggIntDom}
\end{align*}
}%
where $\hat{\bf u}_{1_k}$ is the location of the dominant interferer in $\ccPhi$.
In the following Lemma, we present an expression for $\dE{I_{\tt rem, k}| \hat{D}_{01_k}}{}$.

\begin{lemma}\label{lem:MeanIntDom}
Conditioned on the distance to the dominant interferer $\hat{D}_{01_k}$, the expected interference from the rest of the interfering users is given as 
{\small
\begin{align*}
\dE{I_{\tt rem, k}| \hat{D}_{01_k} = \hat{d}_{01_k}}{} = 2 \pi \int_{\hat{d}_{01_k}}^{\infty} r^{-2\alpha}\ccPhiDen(r, \kappa) r {\rm d}r.\numberthis
\label{eq:MeanIntDom}
\end{align*}
}%
\end{lemma}
\begin{IEEEproof}
Above expression follows from the application of Campbell's theorem.
\end{IEEEproof}

With the knowledge of the expected interference and the distribution of $\hat{D}_{01_k}$, in the following Proposition, we present the coverage probability expression for the CC user of interest.

\begin{prop}\label{prop:CCPc}

Conditioned on the event that the $k$-th pilot is used in the $0$-th cell, the coverage probability of the user that is assigned this sequence is given as
{\small
\begin{equation}
\mathtt{P_{c, k}} (T)= \nbbE_{{D}_{00_k}}\nbbE_{\hat{D}_{01_k}}\bigg[\mathbf{1}\bigg(\hat{d}_{01_k}^{-2\alpha} + \dE{I_{\tt rem, k}| \hat{d}_{01_k}}{} < \frac{{d}_{00_k}^{-2\alpha}}{T} \bigg)\bigg| {\cal I} (0, k) = 1\bigg],
\end{equation}
}%
where $f_{\hat{D}_{01_k}}(\hat{d}_{01_k})$ is given in \eqref{eq:PDFCDFDomIntDisThrm}, and the $\cdf$ of ${D}_{00_k}$ is given in \eqref{eq:Djjk}.
\end{prop}
\begin{IEEEproof}
Conditioned on ${\cal I}(0, k) = 1$, the coverage probability of the user assigned the $k$-th sequence is $\dP{\sinr_{0_k} > T | {\cal I} (0, k) = 1} =$
{\small
\begin{align*}
\dP{\frac{{D}_{00_k}^{-2\alpha}}{T} > I_{\tt agg, k} \bigg| {\cal I} (0, k) = 1}  
= \nbbE_{{D}_{00_k}}\nbbE_{\hat{D}_{01_k}}\bigg[\mathbf{1}\bigg(\hat{d}_{01_k}^{-2\alpha} + \dE{I_{\tt rem, k}| \hat{d}_{01_k}}{} < \frac{{d}_{00_k}^{-2\alpha}}{T} \bigg)\bigg| {\cal I} (0, k) = 1\bigg],
\end{align*}
}%
which completes the proof of the above proposition.
\end{IEEEproof}

\subsection{$\sinr$ coverage analysis of a user assigned to the $l$-th CE pilot}
Most of the intermediate steps necessary for the coverage probability result in this case can be derived on the similar lines as that of the previous section.
Hence, we omit a few of the proofs to avoid repetition.

\subsubsection{Density function of $\Phi_{\tt u, l}$}
\begin{figure}
  \centering
  \includegraphics[width=0.4\columnwidth]{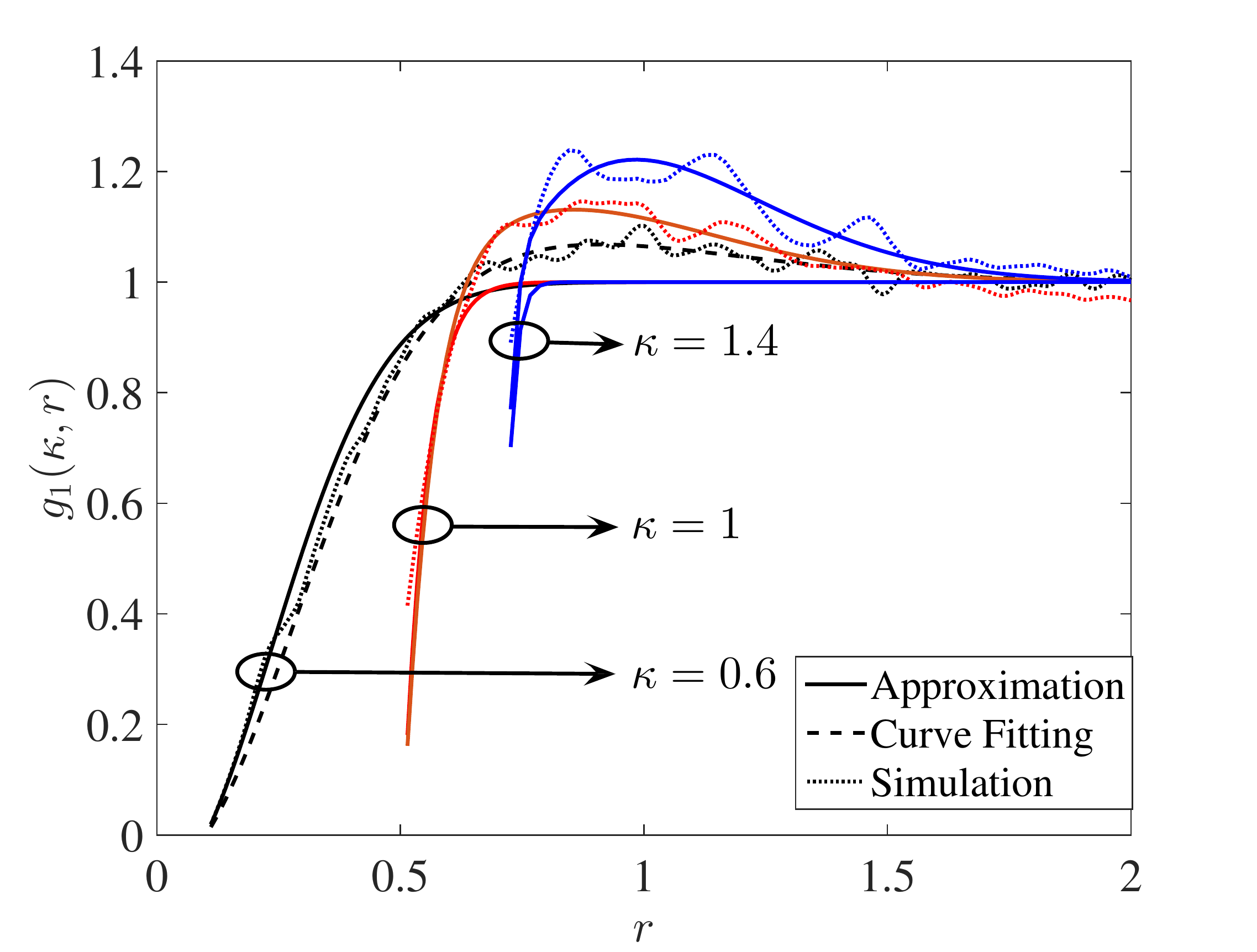}
  \vspace{-0.25cm}
  \caption{\small PCF of $\Phi_{\tt u,CE}$ for different $\kappa$. The approximation and  curve-fitting are based on \eqref{eq:PCF_CE} and \eqref{eq:CurveFitPCF_CE}, respectively.}
    \vspace{-0.25cm}
  \label{fig:PCFCEKappa}
\end{figure}

To begin with, we present the density function of the point process $\Phi_{\tt u, CE}$.
Similar to the CC case, we first present the PCF $g_{\lambda}^{\rm CE}(r, \kappa)$ for $\Phi_{\tt u, CE}$ w.r.t. the $0$-th BS.
Due to scale invariance, we consider a network with unit BS density and threshold radius $\kappa/\sqrt{\pi c_2}$.  
In the following Lemma, we present the expression for $g_{1}^{\rm CE}(r, \kappa)$.
\begin{lemma}\label{lem:PCFCE}
The PCF of $\Phi_{\tt u, CE}$ w.r.t. the $0$-th BS is given as 
{\small
\begin{align}
g_{1}^{\rm CE}(r, \kappa) \approx 1 - \exp\left(- \pi \left(r^2 - \frac{\kappa^2}{\pi c_2} \right) \frac{14}{5} \dP{{\cal E}_3^C} \exp(\kappa^2/c_2) \right), \quad r \geq \frac{\kappa}{\sqrt{\pi c_2}}.
\label{eq:PCF_CE}
\end{align}
}%
\end{lemma}
\begin{IEEEproof}
Please refer to Appendix~\ref{app:PCFCE}.
\end{IEEEproof}
In Fig.~\ref{fig:PCFCEKappa}, we present the PCF for different values of $\kappa$. 
The approximate theoretical expression presented in~\eqref{eq:PCF_CE} is compared with the simulation results. In addition, following prototype function is also used to approximate the PCF for comparison purpose
{\small
\begin{align}
\hat{g}_{1}^{\rm CE}(r, \kappa) = 1 - \exp(-a (r^2 - R_c^2)) + b(r^2 - R_c^2)\exp(-c(r^2 -R_c^2)),
\label{eq:CurveFitPCF_CE}
\end{align}
}%
where the values of the parameters $a, b, c$ are obtained through curve fitting with simulated PCF.
Based on the figure, we make the following remark regarding the accuracy of the PCF in~\eqref{eq:PCF_CE}.
\begin{remark}
As $\kappa$ increases, the PCF obtained from simulation indicates that $\Phi_{\tt u, CE}$ exhibits clustering behaviour beyond $R_c$. 
However, by approximating the PCF using the exponential function presented in~\eqref{eq:PCF_CE}, it is not possible to capture this clustering nature.
More complicated functions such as \eqref{eq:CurveFitPCF_CE} can be used for this purpose. However, determining the values of the parameters $a, b$, and $c$ analytically is not tractable.
Note that from the network deployment perspective higher values of $R_c$ may not be desirable, because it would result in higher fraction of cells without CE region. 
Hence, CE pilot will be underutilized reducing the benefit of FPR.
Therefore, the range of $\kappa$ for which the approximation of PCF using \eqref{eq:PCF_CE} is poor is of lesser practical importance.
\end{remark}

Similar to the CC case, we approximate ${\Phi}_{\tt u, CE}$  as a non-homogeneous PPP with density function
{\small
\begin{align}
\lambda_{\tt u, CE}^{\rm PPP} (r, \kappa) = \lambda_0 \dP{{\cal E}_3^C} \left(1 - \exp(-\pi \lambda_0 (r^2 - R_c^2) \dP{{\cal E}_3^C} \frac{14}{5} \exp(\kappa^2/c_2))\right), \quad r \geq R_c.
\end{align}
}%
Recall that $\Phi_{\tt u, l} \subseteq {\Phi}_{\tt u, CE}$ contains the locations of the interfering CE users that use the $l$-th pilot. Similar to the CC case, we approximate $\Phi_{\tt u, l}$ as a non-homogeneous PPP $\cePhi$ whose density function is presented in the following Lemma.
\begin{lemma}
The density function of the PPP $\cePhi$ containing the locations of the active CE interfering users is given as
{\small
\begin{align*}
\cePhiDen(r, \kappa) \approx \lambda_0 \dE{{\cal I}(j, l)}{} \dP{{\cal E}_3^C} \left(1 - \exp\left(-\pi  \frac{14}{5} \exp(\kappa^2/c_2) \dP{{\cal E}_3^C} \lambda_0 (r^2 - R_c^2)\right)\right), \quad r \geq R_c,  
\end{align*}
}%
and corresponding intensity measure is given as 
$
\cePhiIM(r, \kappa) = 2 \pi \int_{t=0}^r \cePhiDen(t, \kappa) t {\rm d}t.
$
\end{lemma}
\begin{IEEEproof}
The density function is obtained on the similar arguments as that of Lemma~\ref{lem:DenFunCC}.
\end{IEEEproof}

\subsubsection{Coverage probability of the CE user of interest}
Similar to the CC case, we use the dominant interferer based approach to obtain the coverage probability expression. 
Using the intensity measure and density function of $\cePhi$, the $\cdf$ and $\pdf$ of the distance to the dominant CE interferer are given as 
{\small
\begin{align}
F_{\hat{D}_{01_l}}(\hat{d}_{01_l}| \kappa) = 1 - \exp(-\cePhiIM(\hat{d}_{01_l}, \kappa)),\quad f_{\hat{D}_{01_l}}(\hat{d}_{01_l}| \kappa) & = 2 \pi \hat{d}_{01_l} \cePhiDen(\hat{d}_{01_l}, \kappa)\exp(-\cePhiIM(\hat{d}_{01_l},\kappa)).
\label{eq:PDFCDFD01_l}
\end{align}
}%
Now, conditioned on the distance to the dominant interferer $\hat{D}_{01_l}$, the aggregate interference at the $0$-th BS from the CE users is approximated as $I_{\tt agg, l} = $
{\small 
\begin{align*}
\hat{d}_{01_l}^{-2\alpha} + \nbbE\bigg[\sum_{{\hat{\bf u}_{j_l}} \in \cePhi \setminus {\hat{\bf u}_{1_l}}} \hat{d}_{0j_l}^{-2\alpha} \bigg| \hat{d}_{01_l}\bigg] = \hat{d}_{01_l}^{-2 \alpha} +  \dE{I_{\tt rem, l} \bigg|\hat{d}_{01_l}}{} \stackrel{(a)}{=} \hat{d}_{01_l}^{-2\alpha} + 2 \pi \int_{\hat{d}_{01_l}}^{\infty} r^{-2\alpha}\cePhiDen(r, \kappa) r {\rm d}r,\numberthis
\label{eq:AggIntDomCE}
\end{align*}
}%
where $(a)$ follows from the application of Campbell's theorem.
Using the above expression for aggregate interference, the coverage probability of the CE user of interest is presented next.
\begin{prop}\label{prop:CEPc}
Conditioned on the event that ${\cal I}(0, l) = 1$, the coverage probability of a user assigned to this pilot is given as
$\mathtt{P_{c, l}} (T)= \dP{\sinr_{\tt 0, l} > T| {\cal E}_3^C, {\cal I}(0, l) =1} = $
{\small
\begin{align*}
\dP{\frac{D_{00_l}^{-2 \alpha}}{T} > I_{\tt agg, l} \bigg| {\cal E}_3^C, {\cal I}(0, l) =1} = \nbbE_{D_{00_l}} \nbbE_{\hat{D}_{01_l}} \bigg[\mathbf{1}\bigg(\hat{d}_{01_l}^{-2\alpha} + \dE{I_{\tt rem, l} \bigg|\hat{d}_{01_l}}{} < \frac{d_{00_l}^{-2 \alpha}}{T}\bigg)\bigg| {\cal E}_3^C, {\cal I}(0, l)=1\bigg]. \numberthis
\end{align*}
}%
\end{prop}
\begin{IEEEproof}
The proof can be done on the similar lines as that of Proposition~\ref{prop:CCPc}.
\end{IEEEproof}

\subsection{Average user SE and cell SE}
Using the coverage probability results, in the following Proposition, we present the approximate expressions for average SE of the CC and CE users of interest.
\begin{prop}\label{prop:AvgUSE}
The average SE of a randomly selected CC user is given as 
{\small
\begin{align*}
\overline{\se}_{\tt u, CC} \approx & B_C \dE{{\cal A}_{\tt 0k, CC}}{} \int_{t=0}^{\infty}\mathtt{P_{c, k}} (2^t -1) {\rm d}t, \numberthis
\end{align*}
}%
where $\mathtt{P_{c, k}} (\cdot)$ is presented in Proposition~\ref{prop:CCPc} and $\dE{{\cal A}_{\tt 0k, CC}}{}$ is presented in Lemma~\ref{lem:PilotAssgnCC}.
Similarly, the average SE of a randomly selected CE user is given as 
{\small
\begin{align}
\overline{\se}_{\tt u, CE} \approx & B_E \dE{{\cal A}_{\tt 0l, CE}|{\cal E}_3^C}{} \int_{t=0}^{\infty}\mathtt{P_{c, l}} (2^t -1) {\rm d}t. \numberthis
\end{align}
}%
\end{prop}
\begin{IEEEproof}
From \eqref{eq:userSE}, the average SE of the CC user of interest can be approximated as
{\small
\begin{align*}
\overline{\se}_{\tt u, CC} \approx & \dE{\sum_{n=1}^{B_C} {\cal A}_{\tt 0n, CC}\log_2(1 + \sinr_{0_n})}{} = B_C \dE{{\cal A}_{\tt 0k, CC}\log_2(1 + \sinr_{0_k})}{} \approx B_C \dE{{\cal A}_{\tt 0k, CC}}{} \dE{\log_2(1 + \sinr_{0_k})}{},
\end{align*}
}%
where $\sinr_{0_n}$ is the $\sinr$ of the CC user of interest if it is assigned the $n$-th CC pilot. The last step in the above expression follows from the independence assumption between ${\cal A}_{\tt 0k, CC}$ and $\sinr_{0_k}$.
The expression in the lemma follows from the last step using the fact that for a positive random variable \mbox{\small $X$, $\dE{X}{} = \int_0^\infty \dP{X > t} {\rm d}t$}.
The average CE user SE can be derived on the similar lines.
\end{IEEEproof}

Now, we present the expression for average cell SE.
\begin{prop}\label{prop:AvgCSE}
The average cell SE of a typical cell is given as $\overline{\cse} = $
{\small
\begin{align*}
\left(1 - \frac{B}{T_c}\right) \left(B_C \dE{{\cal I}(0, k)}{} \dE{\log_2(1 + \sinr_{0_k})}{} + \dP{{\cal E}_3^C} B_E \dE{{\cal I}(0, l)|{\cal E}_3^C}{} \dE{\log_2(1 + \sinr_{0_l})|{\cal E}_3^C}{}\right). \numberthis
\end{align*}
}%
\end{prop}
\begin{IEEEproof}
From \eqref{eq:SumCellSE}, we write $\dE{\cse}{}$
{\small
\begin{align*}
& \stackrel{(a)}{=} \left(1 - \frac{B}{T_c}\right)\left( \dE{\sum_{n=1}^{B_C} \log_2( 1 + \sinr_{0_n})}{} +  \dP{{\cal E}_3^C} \dE{\sum_{m=1}^{B_E} \log_2( 1 + \sinr_{0_m})\bigg|{\cal E}_3^C}{}\right) \\
& \stackrel{(b)}{=}  \left(1 - \frac{B}{T_c}\right) \left(B_C \dE{{\cal I}(0, k)}{} \dE{\log_2( 1 + \sinr_{0_k})}{} + B_E \dE{{\cal I}(0, l)|{\cal E}_3^C}{} \dP{{\cal E}_3^C}\dE{\log_2( 1 + \sinr_{0_l})|{\cal E}_3^C}{}\right),
\end{align*}
}%
where $(a)$ follows from the law of total probability and $(b)$ follows from the fact that the selection of pilots are equi-probable events.
\end{IEEEproof}

\vspace{-0.5cm}
\section{Numerical Results and Discussion}
In this section, we validate the approximate theoretical results using Monte Carlo simulations. 
Further, we study the effect of different system parameters on the $\sinr$ coverage probability, average user and cell SEs. 
In our simulation framework, we consider the BS density $\lambda_0 = 4 \times 10^{-6}$, pathloss exponent $\alpha = 3.7$,  the coherence time interval $T_c = 200$ symbol duration, and the pilot length $B = 100$ symbol duration. 
For comparison purpose, we also provide SE results corresponding to pilot reuse-1 at necessary places.
{Note that the system model for reuse-1 is the same as described in Sec.~\ref{sec:SysMod}. The key difference is that there is no segregation in terms of CC (CE) pilots and the entire set of $B$ pilots can be assigned to any user attached to a BS. 
This complicates the pilot utilization analysis. 
To be specific, to obtain the probability of the event that a CC (CE) user is assigned a given pilot requires the consideration of the joint distribution of the number of CC and CE users. This result does not directly follow from Lemma~\ref{lem:PilotAssgnCC} and requires additional analysis, which does not appear tractable as deriving joint distribution for the CC and CE areas of a typical cell is challenging. The similar remark holds for the probability of pilot utilization in case of reuse-1.
Hence, to validate the efficacy of FPR scheme with respect to reuse-1, we rely on simulation-based results for reuse-1.}

\vspace{-0.3cm}
\subsection{$\sinr$ coverage probability of a user assigned to a given pilot}
\begin{figure}[!htb]
\centering
\begin{subfigure}{0.32\textwidth}
  \centering
  \includegraphics[width=1\linewidth]{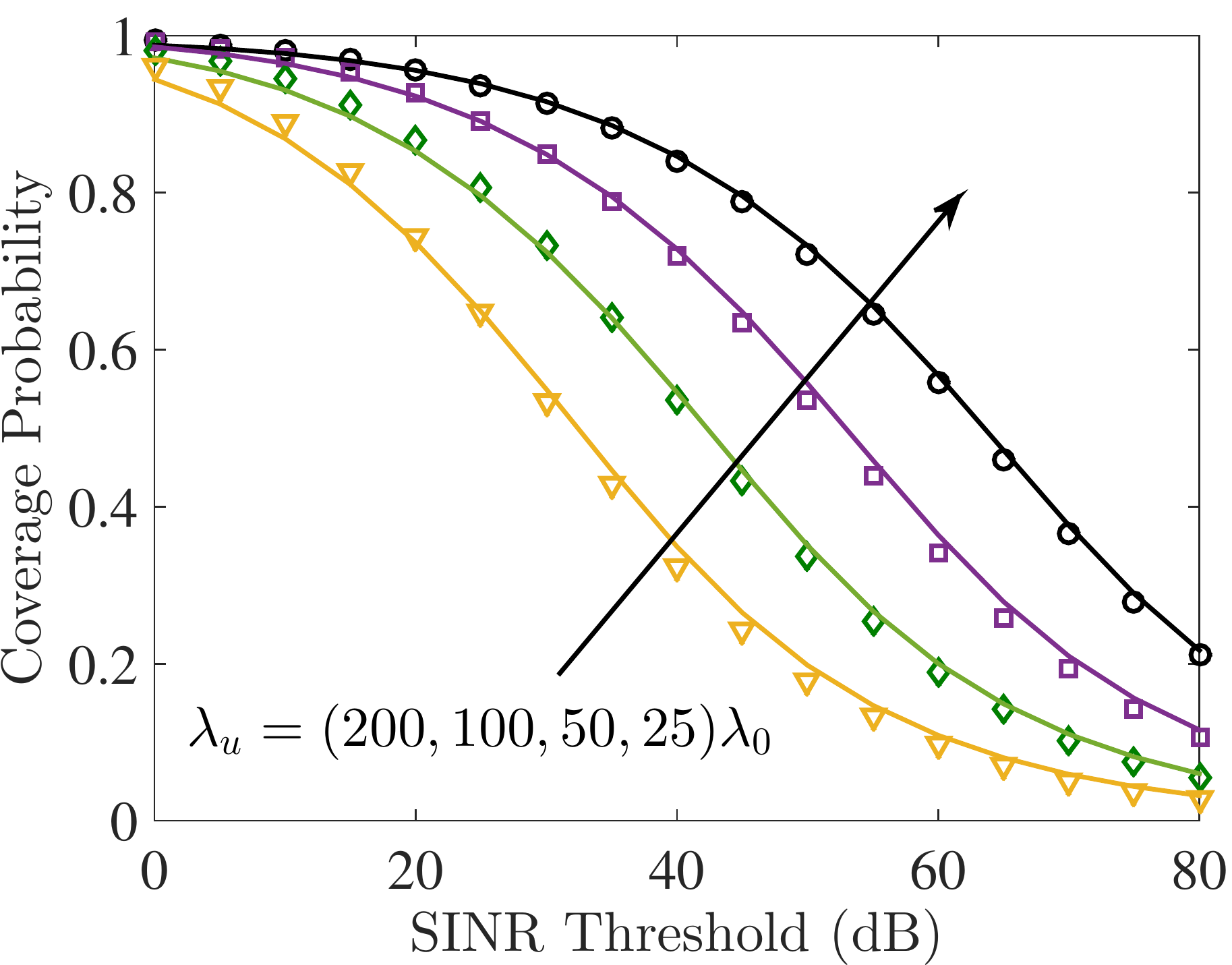}
\end{subfigure}
\hspace{2cm}
\begin{subfigure}{0.32\textwidth}
  \centering
  \includegraphics[width=1\linewidth]{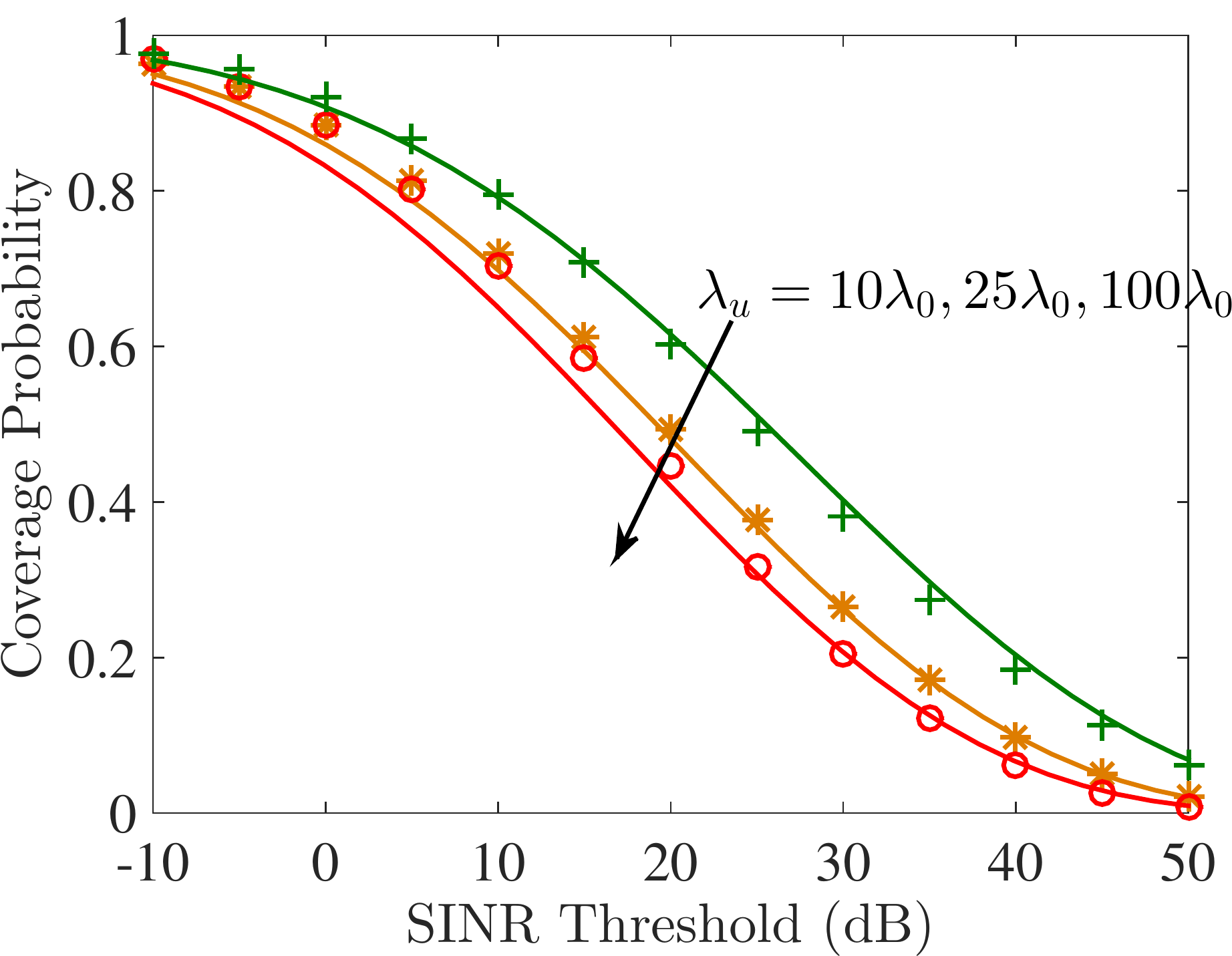}
\end{subfigure}%
\vspace{-0.25cm}
\caption{\footnotesize Coverage probability of a CC user on a given CC pilot (left) and CE user on a given CE pilot (right) for different $\lambda_u$. Markers and solid lines represent the simulation and theoretical results, respectively. $\kappa = 0.6, B_C = 58, B_E = 14, \beta_f = 3$.}
\vspace{-1cm}
\label{fig:PcLamU}
\end{figure}

\begin{figure}[!htb]
\centering
\begin{subfigure}{0.32\textwidth}
  \centering
  \includegraphics[width=1\linewidth]{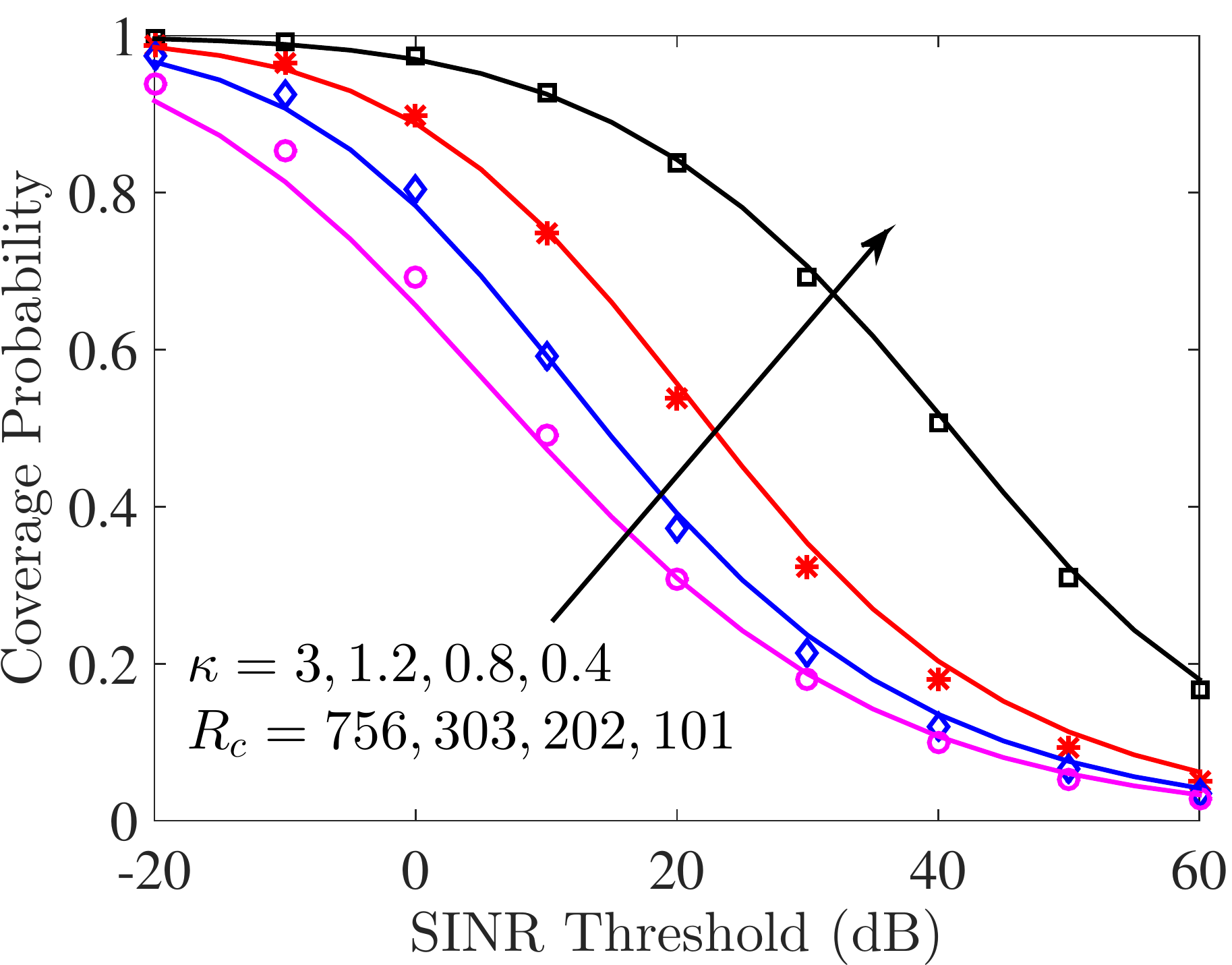}
\end{subfigure}
\hspace{2cm}
\begin{subfigure}{0.32\textwidth}
  \centering
  \includegraphics[width=1\linewidth]{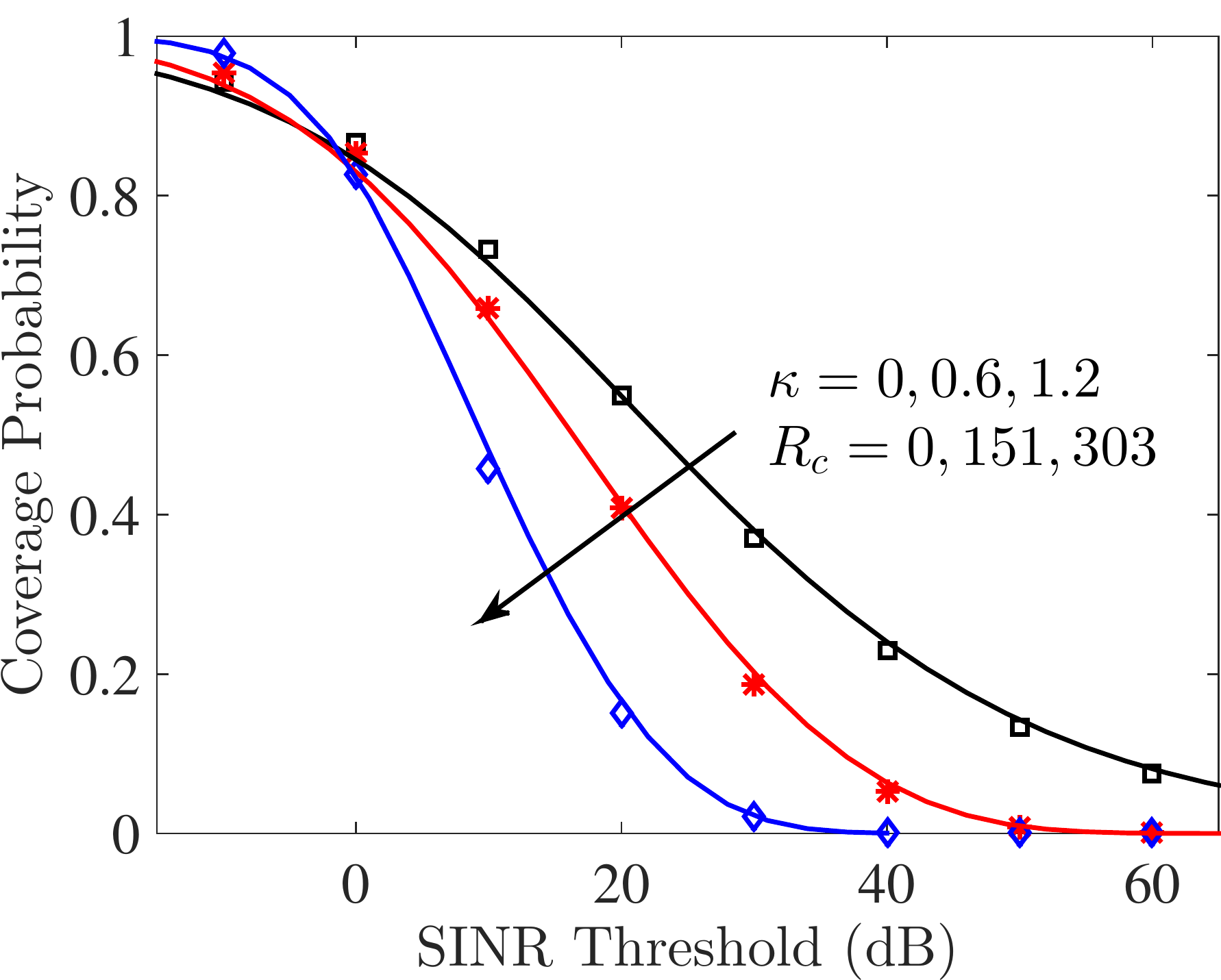}
\end{subfigure}%
\caption{\footnotesize Coverage probability of a CC user on a given CC pilot (left) and CE user on a given CE pilot (right) for different $R_c$. Markers and solid lines represent the simulation and theoretical results, respectively. $B_C = 58, B_E = 14, \beta_f = 3, \dE{{\cal I}(0, k)}{} = \dE{{\cal I}(0, l)|{{\cal E}_3^C}}{} =1$.}
\vspace{-0.25cm}
\label{fig:PcvsKappa}
\end{figure}

In this subsection, we study the effect of different system parameters on the coverage probability of a CC (CE) user that is assigned the $k$-th ($l$-th) pilot.
The effect of $\lambda_u$ on coverage probability for CC and CE cases can be observed from Fig.~\ref{fig:PcLamU} (left and right, respectively).
From the figures, we infer that with the increasing density, the coverage probability reduces in both the scenarios. 
This is intuitive as with increasing $\lambda_u$, the pilot usage probability in the interfering cells increases, thereby increasing the aggregate interference. 
The effect of normalized threshold radius $\kappa$ on coverage probability is presented in Fig.~\ref{fig:PcvsKappa} for CC (left) and CE (right) cases. As observed from Fig.~\ref{fig:PcvsKappa} (left), with decreasing $\kappa$ (equivalently $R_c$), the coverage probability improves. This behavior is justified by the fact that with decreasing $R_c$ the serving distance also decreases. 
In addition, the pilot usage probability in interfering cells also reduces. Combination of both the effects  results in $\sinr$ coverage probability improvement.
For a randomly selected CE user assigned a given CE pilot sequence, above trend is observed for higher $\sinr$ thresholds. On the other hand, for lower $\sinr$ thresholds, reverse trend is observed. One possible explanation behind this behaviour is that although with increasing $R_c$ serving distance increases, the number of interfering users reduces. This  results in improvement of coverage probability.

\vspace{-0.5cm}
\subsection{Average CC (CE) user SE and cell SE}

\begin{figure}[!htb]
\centering
\begin{subfigure}{0.31\textwidth}
  \centering
  \includegraphics[width=1\linewidth]{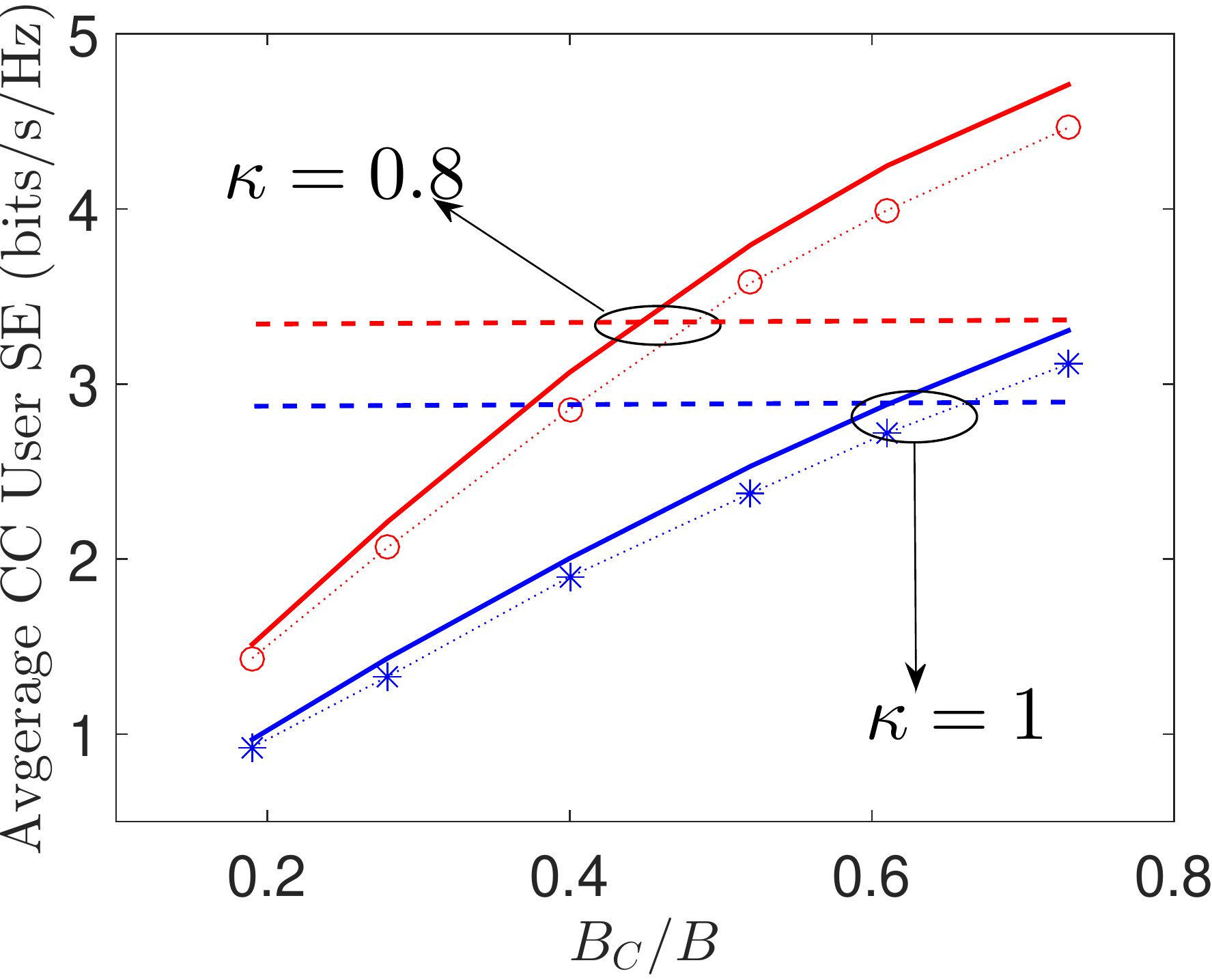}
\end{subfigure}
\begin{subfigure}{0.32\textwidth}
  \centering
  \includegraphics[width=1\linewidth]{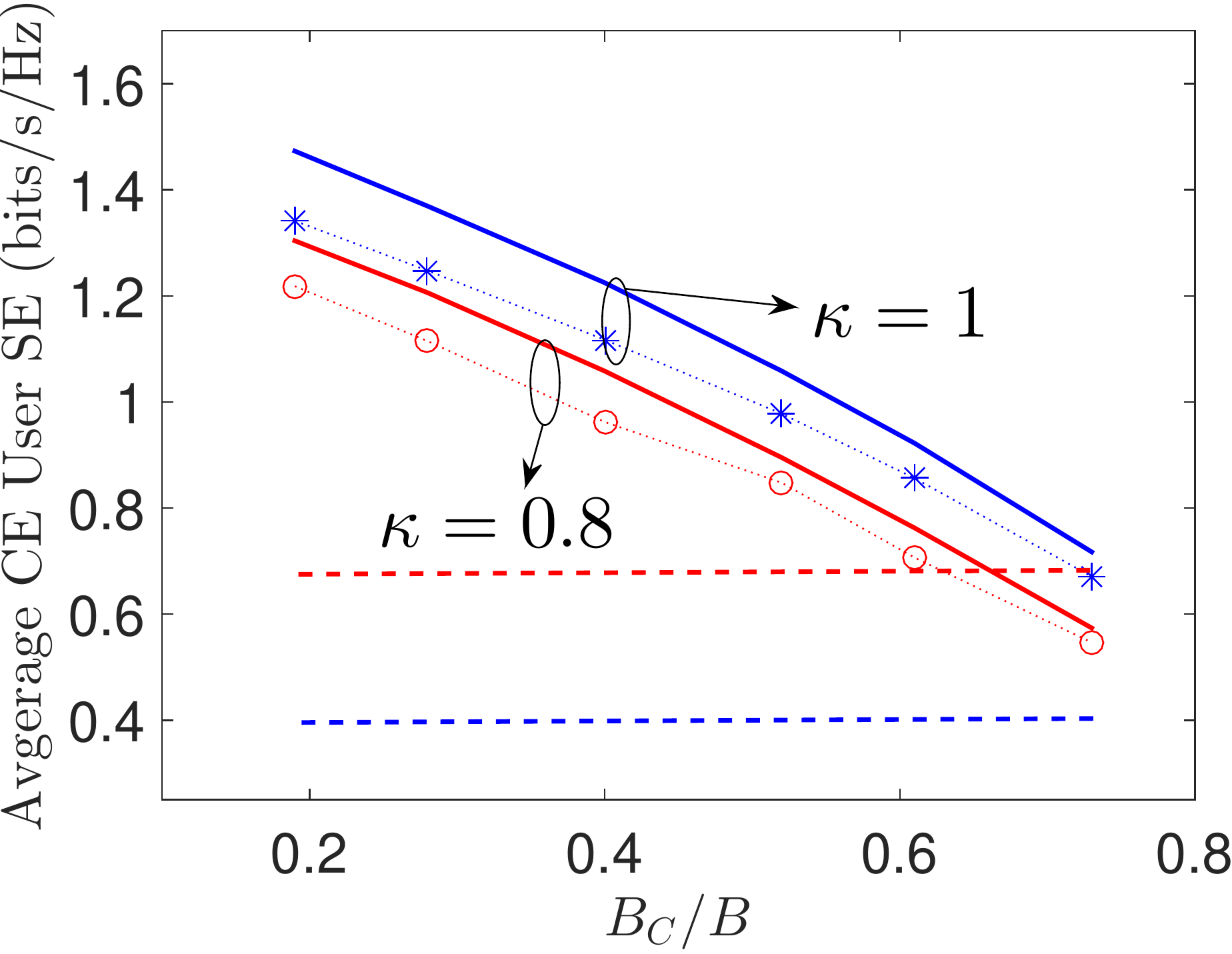}
\end{subfigure}%
\begin{subfigure}{0.32\textwidth}
  \centering
  \includegraphics[width=1\linewidth]{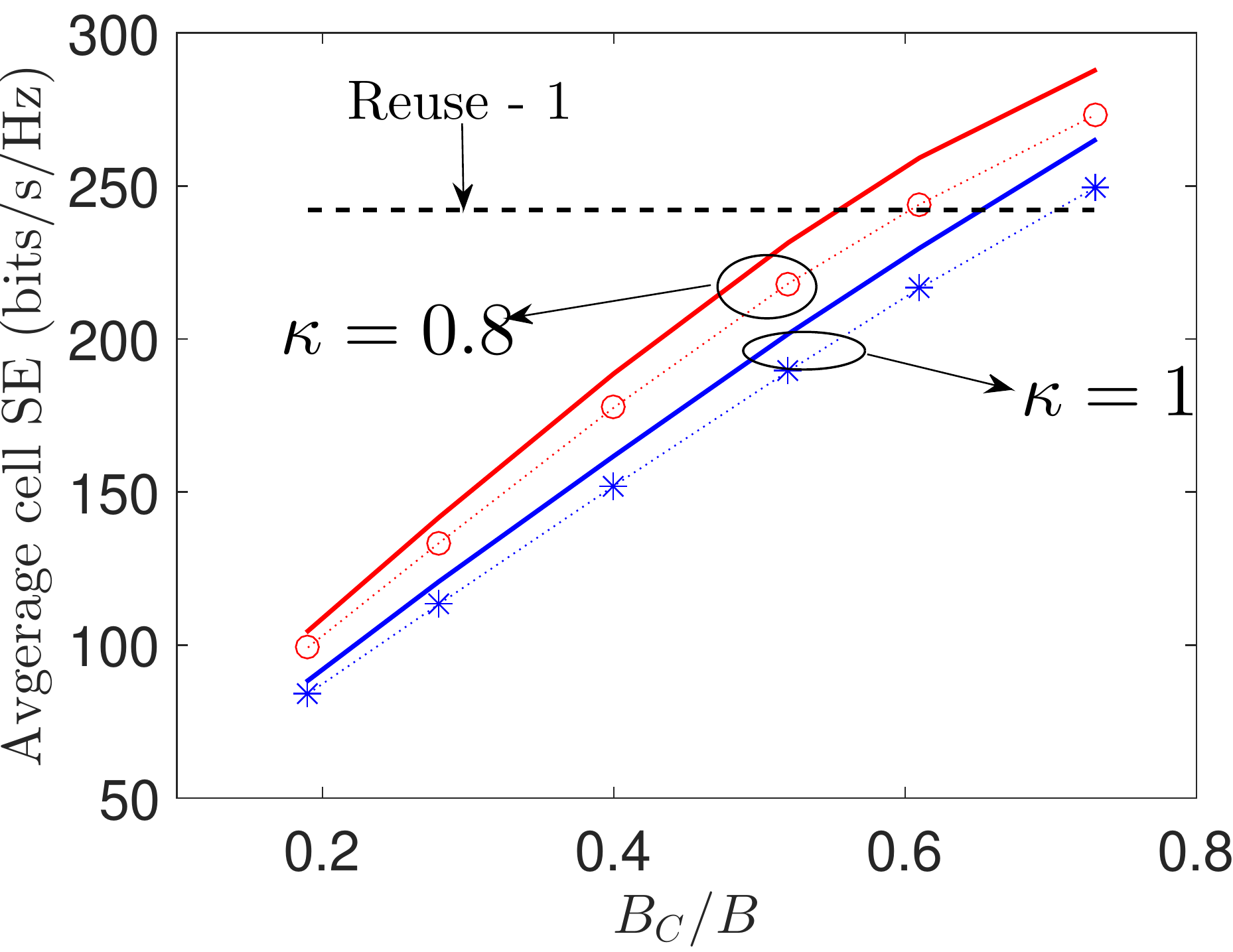}
\end{subfigure}%
\vspace{-0.25cm}
\caption{\footnotesize The average SE of CC user of interest (left), CE user of interest (center), and a typical cell (right) as functions of $B_C/B$. The solid lines and marked dotted lines represent the theoretical and simulation results, respectively. The dashed lines represent the simulated CC user SE corresponding to reuse-1. $B = 100, \lambda_{\tt u} = 150 \lambda_0, \lambda_0 = 4 \times 10^{-6}, \beta_f = 3$.}
\vspace{-0.25cm}
\label{fig:USEvsBcbyB}
\end{figure}

{\bf SE as a function of $B_C/B$:} 
In Fig.~\ref{fig:USEvsBcbyB}, the average SEs of CC and CE users of interest as well as a typical cell are presented for different values of $B_C/B$, where $B = 100$. 
For reference, we have also presented the average CC and CE user SEs for unity pilot reuse. 
From Fig.~\ref{fig:USEvsBcbyB} (left), we observe that FPR scheme performs better compared to unity reuse beyond a certain $B_C/B$. 
For both the curves (corresponding to $\kappa = 0.8, 1$), this value of $B_C/B$ lies in the neighbourhood of  $1 - \exp(-\kappa^2)$.
Intuitively, in case of unity reuse, the probability of assigning a pilot sequence to a CC user is approximately $1 - \exp(-\pi \lambda_0 c_2 R_c^2) =  1 - \exp(-\kappa^2)$. 
Hence, on an average $1 - \exp(-\kappa^2)$ fraction of pilot sequences are assigned to CC users. 
Therefore, by choosing $B_C/B \approx 1 - \exp(-\kappa^2)$ in FPR case, the average SE for CC user of interest becomes close to the SE of a CC user in unity reuse.
On the other hand, from Fig.~\ref{fig:USEvsBcbyB} (middle), we observe that for a wide-range of $B_C/B$ the average SE of CE user of interest in FPR is higher compared to average CE user SE in case of unity reuse.
This result justifies the use of FPR scheme as its main purpose is to improve the performance of CE users. 
Finally, the average cell SE for FPR scheme is presented in Fig.~\ref{fig:USEvsBcbyB} (right) for two different values of $\kappa$.
For comparison purpose, the cell SEs corresponding to reuse-1 is also presented. 
Depending on the value of $\kappa$, for certain values of $B_C/B$, cell SE gains over reuse-1 is possible.  

\begin{figure}[!htb]
\centering
\begin{subfigure}{0.32\textwidth}
  \centering
  \includegraphics[width=1\linewidth]{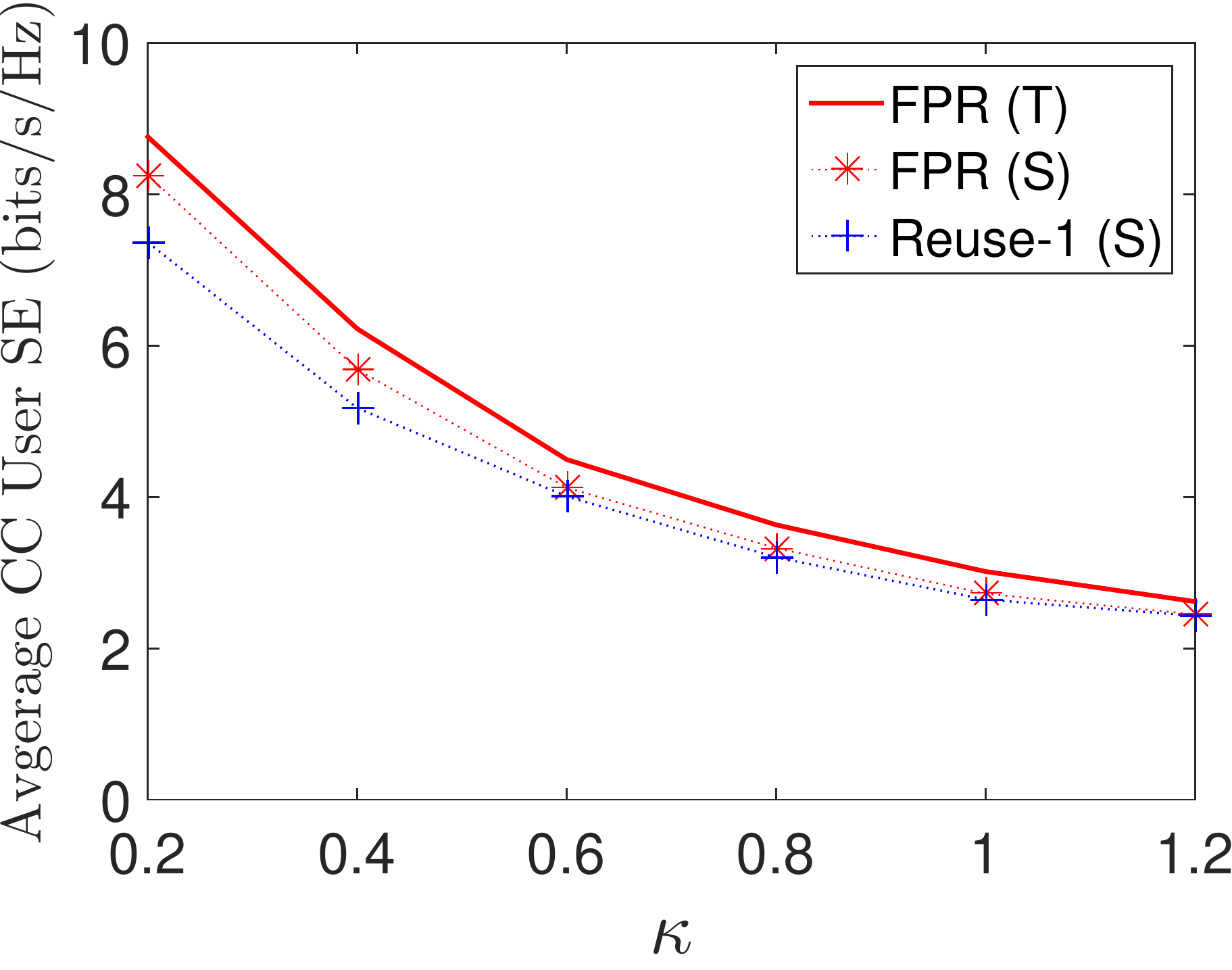}
\end{subfigure}
\begin{subfigure}{0.32\textwidth}
  \centering
  \includegraphics[width=1\linewidth]{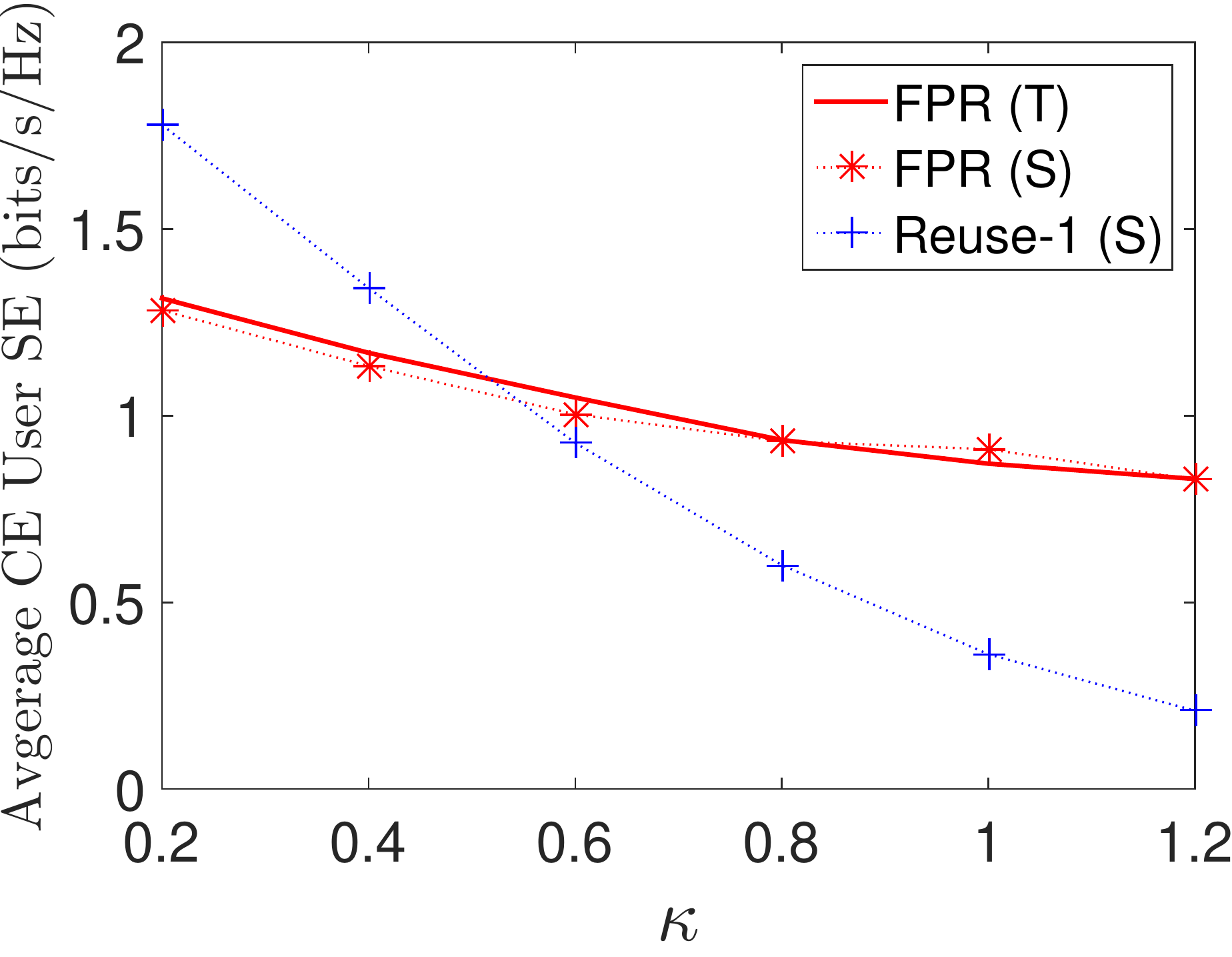}
\end{subfigure}%
\begin{subfigure}{0.32\textwidth}
  \centering
  \includegraphics[width=1\linewidth]{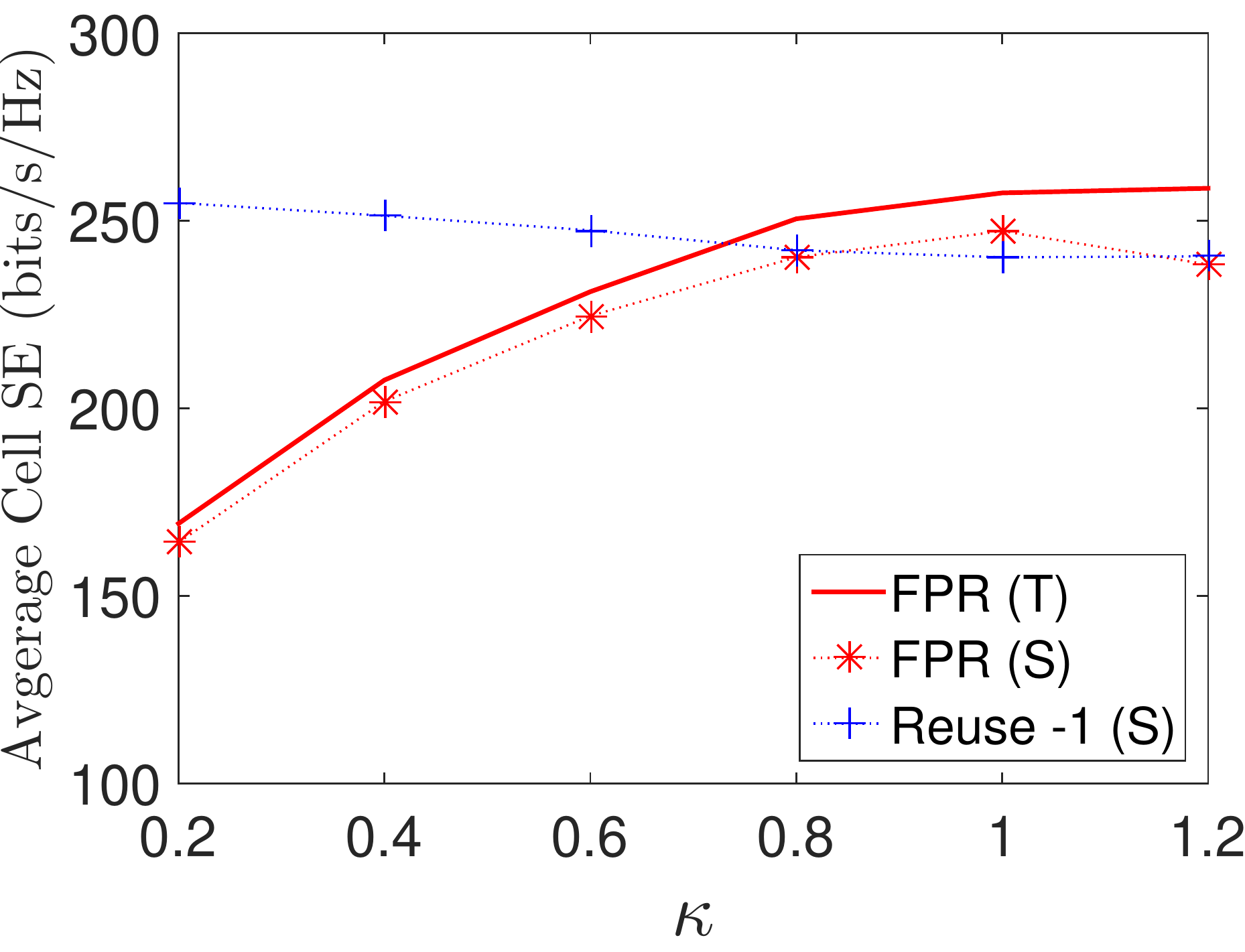}
\end{subfigure}
\vspace{-0.25cm}
\caption{\footnotesize The average SE of CC user of interest (left),  CE user of interest (middle) user, and a typical cell as functions of normalized radius $\kappa$. $\lambda_0 = 4 \times 10^{-6}, \lambda_u = 150 \lambda_0, B_C/B \approx \left(1 -\exp(-\kappa^2)\right), \beta_f = 3$.
Solid lines and marked dotted lines represent theoretical and simulation results, respectively.}
\vspace{-0.25cm}
\label{fig:USEvsKappa}
\end{figure}

{\bf SE as a function of $\kappa$:} The average SEs for the three cases of interest (CC user of interest, CE user of interest, and a typical cell) are presented in Fig.~\ref{fig:USEvsKappa} for different values of $\kappa$.
Based on the insights from the previous result, in order to achieve the same CC user SE as reuse-1, we partition the pilot sequences into two sets such that $B_C/B \approx 1-\exp(-\kappa^2)$.
From Fig.~\ref{fig:USEvsKappa} (left), we observe that aforementioned partitioning rule provides same CC user SE as reuse-1 scheme.
Similarly, in Fig.~\ref{fig:USEvsKappa} (middle), we observe that the CE user spectral efficiency of reuse-1 is better compared to the FPR scheme for lower values of $\kappa$.
This is because of the fact that when $\kappa$ is low, more number of users lie in the CE region. Since FPR employs reuse-3 scheme, the pilot assignment probability to a randomly selected user reduces that results in the reduction of user SE compared to the reuse-1 scheme.
However, for higher values of $\kappa$, FPR performs better compared to the reuse-1 scheme, which is the desired outcome.
From Fig.~\ref{fig:USEvsKappa} (right), we observe that the average cell SE in case of FPR scheme is close to reuse-1 scheme for higher values of $\kappa$ with the above partitioning rule.
System operation at this point is desirable as it improves the CE user SE while providing comparable CC user SE.

\vspace{-0.75cm}
\section{Concluding Remarks}
In this work, we have analyzed the UL performance of a mMIMO system with fractional pilot reuse.
Using tools from stochastic geometry, we have presented approximate expressions for the $\sinr$ coverage probability and average SE of a randomly CC (CE) user in a typical cell.
Our analysis begins with the accurate approximations of the area distributions of CC and CE regions of a typical cell.
These distributions are used to analyze the pilot assignment probability for the user of interest and utilization probability of a given pilot sequence in a typical cell. While the former quantity is directly used in average user SE evaluation, the latter quantity is helpful in obtaining the average cell SE and statistical characterization of interfering user point processes for both CC and CE cases.
All the theoretical results are validated through extensive Monte Carlo simulations.
From our system analysis, we arrive at the conclusion that with proper selection of system parameters it is possible to improve the CE user SE with negligible performance degradation in the CC user SE and cell SE compared to the unity pilot reuse.
There are several possible extensions of this work. 
In this work, we have considered an asymptotically large number of antennas at the BSs. Hence, a natural extension of this work is the consideration of a finite number of antennas.
From stochastic geometry perspective, our analysis of interfering user point process formed by CE users can be improved further by modeling this point process as a cluster process or a Poisson hole process~\cite{Yazda2016}.

\vspace{-0.4cm}
\appendix
\vspace{-0.25cm}

\subsection{Proof of Lemma~\ref{lem:M1M2CE}}\label{app:M1M2CC}
The mean area of the CE region can be expresses as \mbox{\small $\dE{|{\cal X}_E(\nbo, R_c, \Psi_b)|}{} = $}
{\small
\begin{align*}
\dE{\int\limits_{\nbx \in \nbbR^2} \mathbf{1}_{(\nbx \in \vCell_{\Psi_b}(\nbo) \cap {\cal B}_{R_c}^{C} (\nbo))} {\rm d}\nbx}{} \stackrel{(a)}{=} \int_{\nbx \in \nbbR^2 \cap {\cal B}_{R_c}^{C} (\nbo)} \exp(-\pi \lambda_0 \|\nbx\|^2) {\rm d}\nbx = 2 \pi \int_{r=R_c}^{\infty} \exp(-\pi \lambda_0 r^2) r {\rm d}r,
\end{align*}
}%
where $(a)$ follows from that fact that a point located at a distance $\|\nbx\|$ from the origin belongs to $\vCell_{\Psi_b}(\nbo)$, if there are no other BSs in ${\cal B}_{\|\nbx\|}(\nbx)$.
Solving the final integral gives us the expression for the mean in \eqref{eq:M1}.
Similarly, the second moment of the CE area can be expressed as \mbox{\small $\dE{|{\cal X}_E(\nbo, R_c, \Psi_b)|^2}{} = \dE{\int_{\nbx \in \nbbR^2} \mathbf{1}_{(\nbx \in \vCell_{\Psi_b}(\nbo) \cap {\cal B}_{R_c}^{C} (\nbo))} {\rm d}\nbx \int_{\nby \in \nbbR^2} \mathbf{1}_{(\nby \in \vCell_{\Psi_b}(\nbo) \cap {\cal B}_{R_c}^{C} (\nbo))} {\rm d}\nby}{} = $}
{\small
\begin{align*}
& \int_{\nbx \in \nbbR^2} \int_{\nby \in \nbbR^2} \dE{\mathbf{1}_{(\nbx \in \vCell_{\Psi_b}(\nbo)\cap {\cal B}_{R_c}^{C} (\nbo), \nby \in \vCell_{\Psi_b}(\nbo) \cap {\cal B}_{R_c}^{C} (\nbo))}}{} {\rm d}\nbx  {\rm d}\nby \\
\stackrel{(b)}{=} & \int\limits_{(\nbx, \nby) \in \nbbR^2 \cap {\cal B}_{R_c}^{C} (\nbo) \times \nbbR^2 \cap {\cal B}_{R_c}^{C} (\nbo)} e^{-\lambda_0 |{\cal B}_{\|\nbx\|}(\nbx) \cup {\cal B}_{\|\nby\|}(\nby)|} {\rm d}\nbx {\rm d}\nby = 2 \pi \int\limits_{r_1=R_c}^{\infty} \int\limits_{r_2=R_c}^{\infty} \int\limits_{u=0}^{2 \pi} e^{-\lambda_0 V(r_1, r_2, u)} {\rm d}u r_2 {\rm d}r_2  r_1 {\rm d}r_1,
\end{align*}
}%
where $(b)$ follows from the fact that if points $\nbx$ and $\nby$ belong to $\vCell_{\Psi_b}(\nbo)$, then there are no other BSs in the region ${\cal B}_{\|\nbx\|}(\nbx) \cup {\cal B}_{\|\nby\|}(\nby)$, and the last step follows from changing the integration limits from Cartesian to polar coordinates. \hfill
\IEEEQED

\vspace{-0.5cm}
\subsection{Proof of Lemma~\ref{lem:PCFCC}}\label{app:PCFCC}
One approach to deriving $g_1^{\rm CC}(r, \kappa)$ is to first determine the Ripley's K-function $K_1^{\rm CC}(r, \kappa)$ and then use the following relationship:
$
g_1^{\rm CC}(r, \kappa) = \frac{{\rm d}K_1^{\rm CC}(r, \kappa)/{\rm d}r}{2 \pi r}.
$
Note that points in $\Phi_{\tt u, CC}$ are likely to exhibit repulsion w.r.t. $\nbo$ as these points do not lie in $\vCell_{\Psi_b}(\nbo)$.
Since the total interference is likely to be dominated by the nearby users, our main interest lies in characterizing $g_1^{\rm CC}(r, \kappa)$ for small $r$. Note that 
$g_1^{\rm CC}(r, \kappa) \rightarrow 1$ as $r \gg 0$.
Recall that for a point process $\Phi$ of density $\lambda$ the Ripley's K-function is defined as $K_\lambda(r) = \dE{N_{\Phi}({\cal B}_r(\nbo))}{}/\lambda$~\cite{Haenggi2013}, where $N_{\Phi}({\cal B}_r(\nbo))$ denotes the number of points of $\Phi$ that lie in ${\cal B}_r(\nbo)$.
In this case, the K-function is given as 
$
K_1^{\rm CC}(r, \kappa) = \dE{N_{\Phi_{\tt u, CC}}\left(\cup_{\nbx \in \Phi_b}({\cal B}_r(\nbo) \cap \ccReg(\nbx, \kappa/\sqrt{\pi c_2}, \Psi_b))\right)}{}.
$
Now,
{\small
\begin{align*}
K_1^{\rm CC}(r, \kappa) \simeq \dE{N_{\Phi_{\tt u, CC}}\left({\cal B}_r(\nbo) \cap \ccReg(\nby, \kappa/\sqrt{\pi c_2}, \Psi_b)\right)}{}, \quad r \rightarrow 0, \numberthis
\label{eq:K1_1}
\end{align*}
}%
where $\simeq$ denotes approximation that becomes better asymptotically, $\nby$ is the nearest BS to the typical BS at $\nbo$.
Without loss of generality, we assume that $\nby = (\|\nby\|, 0)$.
As per our construction of $\Phi_{\tt u, CC}$, 
we are concerned with only one uniformly distributed point in $\ccReg(\nby, \kappa/\sqrt{\pi c_2}, \Psi_b)$ lying in the region ${\cal B}_r(\nbo) \cap \ccReg(\nby, \kappa/\sqrt{\pi c_2}, \Psi_b)$.
Hence, we write \eqref{eq:K1_1} as
{\small
\begin{align*}
K_1^{\rm CC}(r, \kappa) \simeq \dE{\frac{|{\cal B}_r(\nbo) \cap \ccReg(\nby, \kappa/\sqrt{\pi c_2}, \Psi_b)|}{|\ccReg(\nby, \kappa/\sqrt{\pi c_2}, \Psi_b)|}}{} =  \dE{\frac{S_C(r_m, r, \kappa)}{\ccAre(1, \kappa/\sqrt{\pi c_2})}}{} 
\approx \dE{S_C(r_m, r, \kappa)}{R_m} \dE{\ccAre(1, \kappa/\sqrt{\pi c_2})^{-1}}{},
\end{align*}
}%
where $S_C(r_m, r, \kappa)$ denotes the area of the region ${\cal B}_r(\nbo) \cap {\cal B}_{R_c}(\nby) \cap \left((\nbbR - r_m)^+ \times \nbbR\right)$, and the last approximation follows from independence assumption between $S_C(r_m, r, \kappa)$ and $X_C(1, \kappa/\sqrt{\pi c_2})^{-1}$.
Now, using the result presented in Appendix~\ref{app:ExpArea}, we write
{\small
\begin{equation}
\dE{S_C(r_m, r, \kappa)}{R_m} \simeq
\begin{cases}
\frac{\pi^2 r^4}{2}, & R_c > r, r \rightarrow 0 \\
\pi^2 R_c^2 r^2 - \frac{\pi^2 R_c^4}{2}, &  R_c \leq r, r \rightarrow 0,
\end{cases}
\label{eq:ExpArea}
\end{equation}
}%
where $R_c = \kappa/\sqrt{\pi c_2}$.
The first inverse moment of $\ccAre(1, \kappa/\sqrt{\pi c_2})$ can be evaluated numerically using the approximated distribution presented in Sec.~\ref{sec:AreaDist}. 
Now, the K-function is given as 
{\small
\begin{align*}
K_1^{\rm CC}(r, \kappa) \simeq 
\begin{dcases}
\pi^2 r^4/2 \dE{X_c(1, \kappa/\sqrt{\pi c_2})^{-1}}{} & R_c > r, r \rightarrow 0\\
(\pi^2 R_c^2 r^2 - \pi^2 R_c^4/2) \dE{X_c(1, \kappa/\sqrt{\pi c_2})^{-1}}{} & R_c \leq r, r \rightarrow 0,
\end{dcases}
\end{align*}
}%
and the PCF is given as 
{\small
\begin{align*}
g_1^{\rm CC}(r, \kappa) =\frac{{\rm d}K_1^{\rm CC}(r,\kappa)/{\rm d}r}{2 \pi r} \simeq
\begin{dcases}
\pi r^2 \dE{X_C(1, \kappa/\sqrt{\pi c_2})^{-1}}{} &  R_c > r, r \rightarrow 0 \\
\pi R_c^2 \dE{X_C(1, \kappa/\sqrt{\pi c_2})^{-1}}{} & R_c \leq r, r \rightarrow 0.
\end{dcases}
\end{align*}
}%
Note that as $R_c \rightarrow 0$, the $0$-th BSs observes user locations that are almost identical to BS locations, which is a homogeneous PPP. In this case, when $R_c \rightarrow 0$, $\dE{\ccAre(1, \kappa/\sqrt{\pi c_2})^{-1}}{} \simeq \frac{1}{\pi R_c^2}$. Hence, $g_1^{\rm CC}(r, \kappa) \rightarrow 1$ as expected for a homogeneous PPP. 

Using the asymptotic result that $1 - \exp(-u) \simeq u$ as $u \rightarrow 0$, we write 
{\small
\begin{align*}
g_1^{\rm CC}(r, \kappa) \simeq 
\begin{dcases}
1 -\exp(- \pi r^2 \dE{\ccAre(1, \kappa/\sqrt{\pi c_2})^{-1}}{}) & r < R_c, r \rightarrow 0 \\
1 & r \geq R_c , r \rightarrow 0.
\end{dcases}
\end{align*}
}%
As per the simulation based observation mentioned in \cite{Haenggi2017}, due to the condition $r \rightarrow 0$, the Voronoi cell ${\cal V}_{\Psi_b}(\nby)$ is skewed whose area is likely to be half of the area of a typical Voronoi cell. 
Similar argument can be made for the area of the CC region as well.
Hence, a factor of 2 needs to be introduced for the first condition.
Using this fact, for any value of $r$, a reasonable approximation for the PCF is 
$
g_1^{\rm CC}(r, \kappa) \approx 1 -\exp(- 2 \pi r^2 \dE{\ccAre(1, \kappa/\sqrt{\pi c_2})^{-1}}{}).
$
This completes the proof of the Lemma. \hfill
\IEEEQED

\vspace{-0.5cm}
\subsection{Proof of \eqref{eq:ExpArea}}\label{app:ExpArea}

Depending on the value of \mbox{$R_c$} and \mbox{$r$} we have the following two cases of interest:\newline
{\bf Case~1:} \mbox{$r < R_c$}: The result for this case is obtained from \cite[Lemma~2]{Haenggi2017}, and is given as
{\small
\begin{align*}
\dE{S_C(r_m, r, \kappa)}{R_m} \simeq \frac{\pi^2 r^4}{2} , \quad r \rightarrow 0. 
\end{align*}
}%
{\bf Case~2:} \mbox{$r \geq R_c$}: In this case, the area of the region \mbox{${\cal B}_r(\nbo) \cap \ccReg(\nby, \kappa/\sqrt{\pi c_2}, \Psi_b)$} is given as
{\small
\begin{align*}
S_C(r_m, r, \kappa) =
\begin{dcases}
r^2 \left(u - \frac{\sin{2u}}{2}\right) + R_c^2 \left(v - \frac{\sin{2v}}{2}\right) - (w R_c^2 - r_m \sqrt{R_c^2 - r_m^2}),  & R_c \geq r_m, \\
r^2 u - \frac{r^2}{2}\sin{2u} + R_c^2 v - \frac{R_c^2}{2}\sin{2v}, & R_c < r_m, 
\end{dcases}
\end{align*}
}%
where {\small $R_c = \kappa/\sqrt{\pi c_2}$, $u = \cos^{-1}\left(\frac{r^2 + 4 r_m^2 -R_c^2}{4 r r_m}\right)$, $v = \cos^{-1}\left(\frac{R_c^2 + 4 r_m^2 - r^2}{4 R_c r_m}\right)$, and $w = \cos^{-1}\left(\frac{r_m}{R_c}\right)$.}
Averaging over the random variable $R_m$, we get
{\small
\begin{align*}
\dE{S_C(r_m, r, \kappa)}{} = \pi R_c^2  \int\limits_{0}^{(r-R_c)/2} f_{R_m}(r_m) {\rm d}r_m + \int\limits_{(r-R_c)/2}^{(r+Rc)/2} S(r_m, r, \kappa) f_{R_m}(r_m) {\rm d}r_m,
\end{align*}
}%
where we have used the fact that for $\mbox{\small $r > 2 r_m + R_c$}$, \mbox{\small $S(r_m, r, \kappa) = \pi R_c^2$}.
Further, note that for {\small$2 r_m > r + R_c$,  $S(r_m, r, \kappa) = 0$}. Hence, the upper limit is introduced to consider the values of \mbox{\small $R_m$} for which \mbox{\small $S(r_m, r, \kappa) \neq 0$.}
In addition, we use the asymptotic approximation {\small $f_{R_m}(r_m) =  8 \pi r_m \exp(-4 \pi r_m^2) \simeq 8 \pi r_m (1 - 4 \pi r_m^2)$}, as \mbox{\small $r_m \rightarrow 0$.}
After performing the integration, we obtain \mbox{\small $\dE{S(r_m, r, \kappa)}{} \simeq $}
{\small
\begin{align*}
& \frac{\pi^2 R_c^2 r^4}{2} - \frac{\pi^2 R_c^4 r^2}{2} + \pi^2 R_c^2 r^2 - \left(\frac{\pi^3 R_c^2 r^4}{2} + \frac{\pi^2 R_c^4}{2} + \frac{\pi^3 R_c^6}{2}\right) \simeq  \pi^2 R_c^2 r^2 - \frac{\pi^2 R_c^4}{2}, \quad r \rightarrow 0.
\end{align*}
}%
This completes the proof of \eqref{eq:ExpArea}. \hfill \IEEEQED
\vspace{-0.5cm}
\subsection{Derivation of Lemma~\ref{lem:PCFCE}}\label{app:PCFCE}
The proof can be done on the similar lines as that of Appendices~\ref{app:PCFCC} and \ref{app:ExpArea}.
In this case, the Ripley's K-function is given as 
{\small
\begin{align*}
K_1^{\rm CE}(r, \kappa) \approx \dE{S_E(r_m, r, \kappa)|{\cal E}_3^C}{R_m} \dE{\ceAre(1, \kappa/\sqrt{\pi c_2})^{-1}|{\cal E}_3^C}{}, \quad r \rightarrow 0, r > R_c.
\end{align*}
}%
Asymptotically, conditioned on ${\cal E}_3^C$, the distribution of $R_m$ is given as 
{\small
\begin{align*}
F_{R_m}(r_m | R_M > R_c) = \frac{\dP{R_m \leq r_m, R_M > R_c}}{\dP{R_M > R_c}} \simeq \dP{R_m \leq r_m}, \quad R_c \rightarrow 0.
\end{align*}
}%
The condition $R_c \rightarrow 0$ is of interest to us as our goal is to find the PCF for $r \rightarrow 0$, and  $r > R_c$. 
Now, the following expectation is given as $\dE{S_E(r_m, r, R_c)|{\cal E}_3^C}{R_m} \simeq$
{\small
\begin{align*}
& \int\limits_{0}^{r} A_1(r, r_m, R_c) f_{R_m}(r_m) {\rm d}r_m - \int_{0}^{(r-R_c)/2} A_2(r, r_m, R_c) f_{R_m}(r_m) {\rm d}r_m \\
& - \int\limits_{(r-R_c)/2}^{(r+R_c)/2} A_2(r, r_m, R_c) f_{R_m}(r_m) {\rm d}r_m  - \int\limits_{0}^{R_c} A_3(r, r_m, R_c) f_{R_m}(r_m) {\rm d}r_m \\
= & \frac{\pi^2 r^4}{2} + \frac{\pi^3 r^6}{2} - \frac{\pi^2 R_c^2 r^4}{2} + \frac{\pi^2 R_c^4 r^2}{2} - \pi^2 R_c^2 r^2 + \frac{\pi^3R_c^2r^4}{2} + \frac{\pi^2 R_c^4}{2} + \frac{\pi^3 R_c^6}{2}
\simeq \frac{\pi^2(r^4 + R^4 - 2R_c^2 r^2)}{2},
\numberthis
\end{align*}
}%
where the last step follows from neglecting the $6$-th order terms. 
In the above equation
{\small
\begin{align*}
A_1(r, r_m, R_c) & = r^2 \arccos{\frac{r_m}{r}} - r_m \sqrt{r^2 - r_m^2}, \\
A_2(r, r_m, R_c) & = r^2 u - \frac{r^2}{2}\sin(2u) + R_c^2v - \frac{R_c^2}{2} \sin(2v) \mathbf{1}\left((r-R_c)/2 < r_m \leq (R_c + r)/2\right) + \pi R_c^2 \mathbf{1}\left(r_m \leq (r-R_c)/2\right), \\
A_3(r, r_m, R_c) & = \left(R_c^2 \arccos\left(\frac{r_m}{R_c}\right) - r_m \sqrt{R_c^2 - r_m^2}\right)\mathbf{1}\left(r_m \leq R_c\right).
\end{align*}
}%
Using the above result, the Ripley's K-function is given as 
{\small
\begin{align}
K_1^{\rm CE}(r, \kappa) \simeq \frac{\pi^2}{2} \left(r^2 - R_c^2\right)^2 \dE{X_E(1, \kappa/\sqrt{\pi c_2})^{-1}|{\cal E}_3^C}{}, \quad r > R_c, r \rightarrow 0.
\end{align}
}%
Hence, the PCF is given as 
{\small
\begin{align*}
g_1^{\rm CE}(r, \kappa) =  \frac{{\rm d}K_1^{\rm CE}(r, \kappa)/{\rm d}r}{2 \pi r}  
\simeq  \pi \left(r^2 - R_c^2\right) \dE{X_E(1, \kappa/\sqrt{\pi c_2})^{-1}|{\cal E}_3^C}{}
\approx  \frac{14 \pi \left(r^2 - R_c^2\right) \dP{{\cal E}_3^C}}{5\exp(-\pi R_c^2)},
\end{align*}
}%
where the intuition for the last step follows
from Jensen's inequality $\dE{X_E(1, \kappa/\sqrt{\pi c_2})^{-1}|{\cal E}_3^C}{} \geq (\dE{X_E(1, \kappa/\sqrt{\pi c_2})|{\cal E}_3^C}{})^{-1} = \exp(\pi R_c^2) \dP{{\cal E}_3^C}$. From~\cite{Haenggi2017}, when $R_c = 0, \dE{X_E(1, \kappa/\sqrt{\pi c_2})^{-1}}{} \approx 14/5$. Hence, for $R_c \rightarrow 0$, we approximate $\dE{X_E(1, \kappa/\sqrt{\pi c_2})^{-1}|{\cal E}_3^C}{} \approx 14/5  \exp(\pi R_c^2)\dP{{\cal E}_3^C} = 14/5 \exp(\kappa^2/c_2) \dP{{\cal E}_3^C}$. \hfill
\IEEEQED
\vspace{-0.5cm}
{
\bibliographystyle{IEEEtran}
\bibliography{MasterBibFile}
}%

\end{document}